\documentclass[aps,prd,twocolumn,groupedaddress,nofootinbib,amsmath, amssymb]{revtex4-1}

\usepackage{graphicx}

\begin{document}
	
	
	\title{Physics in precision-dependent normal neighborhoods
	}
	
	
	
\author{Bruno Hoegl}
\email{Bruno.Hoegl@physik.uni-muenchen.de}
\affiliation{Arnold Sommerfeld Center for Theoretical Physics, Theresienstra{\ss}e 37, 80333 M\"unchen, Germany\\}

\author{Stefan Hofmann}
\email{Stefan.Hofmann@physik.uni-muenchen.de}
\affiliation{Arnold Sommerfeld Center for Theoretical Physics, Theresienstra{\ss}e 37, 80333 M\"unchen, Germany\\}

\author{Maximilian Koegler}
\email{M.Koegler@physik.uni-muenchen.de}
\affiliation{Arnold Sommerfeld Center for Theoretical Physics, Theresienstra{\ss}e 37, 80333 M\"unchen, Germany\\}
	%
	
	
	
\date{\today}

\begin{abstract}
We introduce a procedure to determine the size and shape of normal neighborhoods in any spacetimes	and their dependence on the precision of the measurements performed by arbitrary observers. As an example, we consider the Schwarzschild geometry in Riemann and Fermi normal coordinates and determine the size and shape of normal neighborhoods in the vicinity of the event horizon. Depending on the observers, normal neighborhoods extend to the event horizon and even beyond into the black hole interior. It is shown that the causal structure supported by normal neighborhoods across an event horizon is consistent with general relativity. In particular, normal neighborhoods reaching over an event horizon are void of the Schwarzschild coordinate singularity.	In addition, we introduce a new variant of normal coordinates which we call Fermi normal coordinates around a point, unifying features of Riemann and Fermi normal coordinates, and analyze their neighborhoods.
	

\end{abstract}
	
	
	\maketitle
	
	
\section{Introduction}
\label{sec: intro}

Experiments and observations are based on measurement processes with a desired accuracy that depend only to a certain extent on the spacetime geometry. Normal neighborhoods are associated with normal coordinate systems, which allow one to accommodate just the right amount of geometrical data to describe observables with a given accuracy, provided the system under investigation fits into such a neighborhood. Therefore, they enjoy widespread use in many fields of physics. 

For instance, tidal disruption events taking place when stars pass nearby black holes are conveniently 
described in normal neighborhoods. As the tidal forces disrupt the star and strip gas from it, bright and characteristic flares are emitted \cite{TidalBH1,TidalBH2,TidalBH3}, which can be used to detect and characterize the corresponding black hole.

The polynomial nature of normal coordinates allows for a systematic description of physical processes in curved spacetimes.
In particular, using normal coordinates, the geometrical content encoded in the dynamical system under investigation
can be locally approximated, granting approximate solutions to differential equations which cannot be solved on the exact
spacetime geometry.

One has to keep in mind, however, that just as the weak-field approximation is only a local approximation of spacetime, this is similarly (almost) always the case for normal coordinates. Since the infinite normal coordinate expansions usually have to be truncated at some finite order, their validity is restricted to a finite spacetime patch. Therefore, whenever experimental or observational data of a physical system with a given size is to be computed in normal coordinate systems, it is crucial to know their domain of validity and whether the physical system fits into this domain given a desired accuracy. This is also an important question for describing the aforementioned tidal disruption events, which can be seen from \cite{TidalBH1} stating so directly: ``Since the size of the FNC [Fermi normal coordinate] domain is necessarily limited, there is a limit on how long a disrupted star or stripped gas can be followed'' (FNCs are a special choice of normal coordinate systems). For such systems to be describable in normal coordinates, it is obviously required that the normal neighborhood encompasses the disrupted star (and possibly also the star debris). Unfortunately, as of yet this question concerning normal coordinate patch sizes has not been answered satisfactorily in the literature.

The size of normal neighborhoods has until now only been estimated based on curvature arguments, see, for example, \cite{Manasse:1963zz, Domain2, Nesterov:1999ix, Domain4}. Such an estimate is sufficient for calculations aiming for a proof of concept, i.e., when the physical system can always be chosen sufficiently small as, for instance, in \cite{bunchparker,PhysRevD.22.1922,PhysRevD.95.084029,PhysRevD.95.046003}. However, for generic experiments or observations, and as will be seen in the example of tidal disruption events, this estimate is insufficient. Nevertheless, concrete and quantitative calculations concerning the domain of validity of normal coordinate systems and the error that arises from truncating the infinite expansions have not been a focus of discussion in the past. This is the main motivation for this article: 
We show how the shape and size of a normal neighborhood in any spacetime geometry can be calculated explicitly. For that purpose, we consider all observables of interest together with a precision specification, i.e., given a lower resolution bound on the experimental or observational data, we neglect all curvature contributions below the chosen sensitivity. This will restrict the normal coordinates to some spacetime patch of finite size. The spacetime metric is considered as a geometric building block in the construction of observables. The \textit{Mathematica} code we wrote to calculate the patch sizes for this article is provided at~\cite{hoegl_bruno_2020_3966891}. 

An interesting class of spacetimes is the one that contains horizons and singularities. Therefore, we exemplarily apply our method to the geometry of  Schwarzschild black holes and discuss how the resulting patch sizes are seen by arbitrary observers in their corresponding coordinate systems. Additionally, we examine causality in normal coordinate patches encompassing the event horizon.

For the example of tidal interactions as discussed in \cite{TidalBH1}, we determine whether the complete star can be included in a normal neighborhood, given the mass and radius of the star and the black hole in question, as well as their relative distance. 
This example turns out to be quite instructive as it demands a careful calculation of shape and size of the normal neighborhood 
in accordance with an external precision requirement. This will be elaborated at the end of Sec.~\ref{sec: conclusion}. 

A further important topic within the range of black hole tidal interactions are astrophysical jets, especially those of galactic nuclei where the central body is suspected to be a supermassive, rotating black hole. As, for example, in \cite{TidalJets}, the effects of the black hole's tidal forces on the jet particles are used to characterize the black hole.
The effects of gravitational waves on systems such as LIGO can also be calculated using normal coordinates. Although the perturbation of Minkowski spacetime is small, the actual high-precision experiment seems not to fit in a normal neighborhood according to naive size estimates \cite{Nesterov:1999ix}.
Normal coordinates are also employed outside astrophysics and general relativity. An example from biophysics/statistical mechanics is given in \cite{Brownian}, where normal coordinates are employed to describe diffusion processes on the curved manifolds of cell membranes.
A vast subject on its own concerns applications in gauge theories with external fields using the celebrated Fock-Schwinger gauge. This gauge is the analogue to normal coordinates in the sense that it uses Taylor expansions to approximate the gauge field \cite{Muller:1997zk}. The very same procedure presented in this article for normal coordinates can be used to find the domain of validity of the Fock-Schwinger gauge.

We also propose a new variant of Fermi normal coordinates that can be used 
if solving the geodesic equation for the central geodesic in the exact geometry is not possible. In this case, the central geodesic can be computed in terms of a Taylor expanded metric. The result will be a temporally Taylor expanded geodesic that can then be used to set up FNC as usual. These ``FNC around a point'' (FNCP) only require geometrical information at a point, as opposed to the usual ``FNC along a geodesic.''

This article is organized as follows: Sec.~\ref{sec: prelim} contains a short summary of the key aspects of Riemann normal coordinates (RNC). In Sec.~\ref{sec: method} we then present our method for finding the patch size of RNC neighborhoods. Following this, we show sample calculations for the patch size in Sec.~\ref{sec: RNC} and also establish the connection between our result and the familiar patch size estimate presented, for example, in \cite{Domain2} or \cite{Nesterov:1999ix}. In Sec.~\ref{sec: FNC} we discuss FNC and FNCP as well as their domains of validity. In Sec.~\ref{sec: BlackHole} we then compute RNC patches in the geometry of a Schwarzschild black hole, discuss the dependence of those patches on the observer with the Schwarzschild and Painlev\'e-Gullstrand observer as explicit examples. For the latter we also analyze the causal structure in patches ranging across the horizon. Finally, in Sec.~\ref{sec: conclusion} we give our conclusion.

Conventions: Throughout this article, global coordinates $x^a$ assigned to the spacetime and their indices will be denoted by Roman letters, while normal coordinates $\xi^\alpha$ and their indices will be written as Greek letters. Furthermore, whenever an $x$ or $\xi$ dependence of a tensor is not explicitly denoted, the tensor is to be understood as evaluated at the normal coordinate expansion point ($\xi=0$). We choose (anti)symmetrization of $n$ indices to be defined without the $1/n!$ prefactors. Also, we always parametrize curves affinely using their proper length $\tau$ and choose as a parameter for null curves the eigentime of the observer in question. Finally, we choose the signature diag$(-,+,+,+)$, Planck units with $c=G=1$, and the convention for the Riemann curvature tensor $R^a{}_{bcd} = \Gamma^a{}_{b[d,c]} + \Gamma^i{}_{b[d}\Gamma^a{}_{c]i}$.

\section{Preliminaries}
\label{sec: prelim}

According to the principle of relativity, the laws of physics are independent of the observer describing an experiment if the experiment moves uniformly with respect to this observer \cite{Einstein:1905:EBK}. For accelerating experiments one can use the equivalence principle which states that gravitational and inertial mass are equal such that an observer cannot differentiate whether the experiment is accelerating or placed in a homogeneous gravitational field \cite{Einstein:1907:RDGb}.

For objects with finite size both principles apply only if the whole object moves or accelerates uniformly. Nevertheless, this always holds for pointlike objects which allows us to locally rewrite the effects of an inhomogeneous gravitational field as an acceleration \cite{Einstein:1907:RDGb}.

Therefore, in a small neighborhood in which the gravitational field and thus the metric are sufficiently homogeneous, the metric can be brought into Minkowski form by choosing the coordinates in which a freely falling observer is at rest. Normal coordinate systems possess this property. For larger neighborhoods in which the Minkowski metric is insufficient, one can approximate the inhomogeneity of the gravitational field with a Taylor expansion. This results in correction terms to the Minkowski metric. Usually, one has to truncate this infinite expansion after some order, which will in turn result in a mismatch of the approximated metric, compared to the full one, that increases with distance from the reference point. In order for this truncated metric to still be a viable description of the background, it therefore has to be restricted to a domain of validity of finite extent where the mismatch is negligibly small.
	
One coordinate manifestation of the above procedure are Riemann normal coordinates, which are constructed via what is called the exponential map. We will give a summary of their construction following \cite{Petrov, kobayashi_nomitzu}. For a point $p$ on the connected, smooth manifold $M$, let $\gamma_v(\tau)$ with $\gamma_v(0)=p$ and $\left.\mathrm d_\tau\gamma_v\right|_0 = v$ be the geodesics that ``pass through $p$ with velocity $v$.'' Following some $\gamma_v$ for a fixed yet arbitrary length $\tau_0$, the point $\gamma_v(\tau_0)$ is reached and receives, by the exponential map, the coordinates $v$. The exponential map (at $p$) is thus defined as $\text{exp}_p:\, T_pM \rightarrow M$, $v\mapsto \gamma_v(\tau_0)$ with $T_pM$ the tangent space on $M$ at $p$. The rescaling property of geodesics $\gamma_v(a\tau) = \gamma_{av}(\tau)$, $a\in\mathbb{R}$, ensures that the particular choice of $\tau_0$ is irrelevant. Larger $\tau_0$ only exclude some $v$ in $T_pM$, but the same region around $p$ in $M$ is covered by the exponential map.

Subsequently, the $\gamma_v$ are Taylor expanded around $p$, i.e., $\tau=0$, which gives
\begin{eqnarray}
\label{geod_tay_exp}
\!\!\!\!\!\! \gamma^a_v(\tau) &=& p^a + v^a \tau - \frac{1}{2} \Gamma^a{}_{mn} v^m v^n \tau^2  \nonumber \\
\!\!\!\!\!\! && -\frac{1}{6}(\Gamma^a{}_{mn,r} - 2\Gamma^a{}_{ms}\Gamma^s{}_{nr}) v^m v^n v^r\tau^3 + \cdots\;,
\end{eqnarray}
where the geodesic equation was used once for $1/2\left.\mathrm d^2_\tau \gamma\right|_p$ and twice for $1/6 \left.\mathrm d^3_\tau \gamma\right|_p$. In general, one uses the geodesic equation $n-1$ times for the order $n$ coefficients $1/n!\left.\mathrm d^n_\tau \gamma\right|_p$. Now the 4-velocity is expanded in arbitrary vectors $v^a = \lambda^\alpha \gamma^a{}_{,\alpha} = \lambda^\alpha e^a_\alpha$ with $\lambda^\alpha \! \in \mathbb R^{(1,3)}$ the transformed velocity components and $e^a_\alpha$ the vierbein at $p$. The RNC $\lbrace \xi^\alpha \rbrace$ are now introduced as $\xi^\alpha(\tau) = \tau\lambda^\alpha$. Consequently, the RNC coordinate transformation induced by (\ref{geod_tay_exp}) takes the form of a series:
\begin{eqnarray}
\label{RNC_coord_trafo}
\gamma^a_v(\xi) &=& p^a + \xi^\alpha e^a_\alpha - \frac{1}{2} \Gamma^a{}_{mn} e^m_\mu e^n_\nu\xi^\mu\xi^\nu - \frac{1}{6}\left(\Gamma^a{}_{mn,r}  \right. \nonumber \\
&& - \left. 2\Gamma^a{}_{ms}\Gamma^s{}_{nr}\right) e^m_\mu e^n_\nu e^r_\varrho \xi^\mu\xi^\nu\xi^\varrho + \cdots\; .
\end{eqnarray}
Note that the geodesics $\gamma_v$ get mapped onto straight lines $\tau(\lambda^0,\lambda^1,\lambda^2,\lambda^3 )$ in RNC by the exponential map. Furthermore, due to the normalization of $v$ and the orthogonality of the vierbein $g_{ab} e^a_\alpha e^b_\beta = \eta_{\alpha\beta}$, the $\lambda^\alpha$ are normalized with respect to the Minkowski metric $1= g_{ab} v^a v^b = g_{ab} e^a_\alpha e^b_\beta \lambda^\alpha \lambda^\beta = \eta_{\alpha\beta}\lambda^\alpha\lambda^\beta$.

The coordinate transformation (\ref{RNC_coord_trafo}) gives rise to many RNC-specific geometrical identities. Important examples are the two types of identities concerning the $n$th derivatives of the Christoffel symbol:
\begin{align}
\label{geom_id1}
&&\text{(i) }\,& \partial_{(\mu_1}\cdots\partial_{\mu_m}\Gamma^\alpha{}_{\beta\gamma)} = 0 \, ,\, m\in\mathbb{N}_0 \;,\\
\label{geom_id2} \tag{4a}
&&\text{(ii) }\,& \partial_{(\mu_1}\Gamma^\alpha{}_{\beta)\gamma} = \frac{1}{3}R^\alpha{}_{(\beta\mu_1)\gamma} \; , \\
\label{geom_id3} \tag{4b}
&&& \partial_{(\mu_1}\partial_{\mu_2}\Gamma^\alpha{}_{\beta)\gamma} = \frac{1}{2}\partial_{(\mu_1}R^\alpha{}_{\mu_2\beta)\gamma}\; , \,\ldots\;.
\end{align}
\addtocounter{equation}{1}
The $m=0$ case in (\ref{geom_id1}) yields the vanishing Christoffel symbol at the origin. The infinitely many relations of the second type will be denoted as $\lbrace(4)\rbrace$. These identities can then be used to determine the coefficients of a metric Taylor expansion in $\xi$. Up to the so-called adiabatic order~3 of the expansion in $\xi$, the well-known RNC metric series reads
\begin{equation}
\label{RNC_met_intro}
g^{(3)}_{\alpha\beta}(\xi) = \eta_{\alpha\beta} - \frac 1 3 R_{\alpha\mu\beta\nu} \xi^\mu \xi^\nu - \frac{1}{6} R_{\alpha\mu\beta\nu,\varrho} \xi^\mu \xi^\nu \xi^\varrho\;,
\end{equation}
where we denoted the adiabatic order of the truncated metric series in the superscript.

Another way of constructing normal coordinates are FNC developed in \cite{Manasse:1963zz}, where one requires geometrical information along (some interval of) a geodesic. This geodesic then serves as a collection of reference points such that RNC can be set up in the orthogonal directions at every point of the curve. FNC can therefore take into account a preferred curve of a physical system given by this central geodesic.

In theory, the metric of RNC and FNC can be found up to arbitrary order by Taylor expanding and using the identities (\ref{geom_id1}) and $\lbrace(4)\rbrace$, but for calculational purposes one usually has to truncate the series expansion after some order. This will, as discussed above, restrict the series's validity to some limited spacetime patch. For example, when truncating the RNC metric (\ref{RNC_met_intro}) after adiabatic order 2, we require the third term to be negligibly small compared to the first two. This then restricts the possible values of $\xi$, resulting in a finite patch size.

\section{Method for Finding the Patch Size of RNC Neighborhoods}
\label{sec: method}

We will now give a short description how to generally determine the size of an RNC neighborhood. After this, we will discuss each step in greater detail. The following procedure introduces the steps necessary to calculate concrete RNC patch sizes in which the metric and all other tensors of interest are valid given a quantified precision requirement:
\vspace{2mm}

\noindent\textbf{\textit{Step 1}}\\
\indent Use the geometric identities (\ref{geom_id1}) and $\lbrace(4)\rbrace$ to calculate the RNC metric $g(\xi)$ up to adiabatic order $n+1$ in $\xi$ with $n$ being the desired order.
\vspace{1mm}

\noindent\textbf{\textit{Step 2}}\\
\indent Require the $n+1$ order terms to be negligible compared to the metric $g(\xi)$ truncated after the order $n$ denoted by $g^{(n)}(\xi)$. This demand restricts the RNC patch.
\vspace{1mm}

\noindent\textbf{\textit{Step 3}}\\
\indent Build all tensors $T(g^{(n)}(\xi))$ of interest (e.g., the Riemann curvature tensor) using the metric expansion up to order $n$.
\vspace{1mm}

\noindent\textbf{\textit{Step 4}}\\
\indent Demand the calculated $T(g^{(n)}(\xi))$ coincide with their usual Taylor expansions in $\xi$, thus introducing additional conditions. Take the most restrictive condition of these together with Step~2 to determine the patch size.
\vspace{1mm}

\noindent\textbf{\textit{Step 5}}\\
\indent Compute the patch size along a geodesic $\gamma_v$ as seen by the observer corresponding to some $x$ coordinate system by finding the straight line in RNC corresponding to $\gamma_v$ and determining the line's proper length using the RNC conditions from Steps~2 and 4. Then reparametrize $\gamma_v$ with the observer's eigentime and plug in the maximal eigentime determined by the maximal proper length.

\begin{center}
\textbf{\textit{Comments}}
\end{center}

In the following we discuss extensively each of the above steps and the details of our procedure. In Sec.~\ref{subsec: RNC_applied} we then show as an example how the procedure can be applied to determine the patch size for $n=2$ and the Riemann tensor being the additional tensor of interest.

\mbox{}

\noindent\textbf{\textit{Step 1}}\\
\indent The determined $g^{(n)}$~patch size becomes more accurate the more higher order terms are computed and used for the later comparison with $g^{(n)}$. Perfect accuracy is therefore achieved when comparing $g^{(\infty)}-g^{(n)}$ with $g^{(n)}$. For most applications, however, using the order $n+1$ term of the metric $g(\xi)$ denoted by $\mathcal{O}_g(\xi^{n+1})$ is sufficient. Important examples where this is insufficient are Minkowski patches in close vicinity to a black hole of any mass. In this case it is mandatory to calculate the metric at least up to order $3$. We will discuss this in detail in Sec.~\ref{subsec: PG}.\\

\noindent\textbf{\textit{Step 2}}\\
\indent This restriction is always necessary when the infinite metric expansion is inaccessible and a truncation has to be performed. The smallness demand for every metric component then reads
\begin{equation}
\label{2)_general}
\left|\mathcal{O}_{g_{\alpha\beta}}(\xi^{n+1})\right| \leq \varepsilon\, \left|g^{(n)}_{\alpha\beta}(\xi)\right|\;,
\end{equation}
where a smallness-parameter $\varepsilon \in \;]\,0,1]\,$, usually $\varepsilon\ll 1$, was introduced to encode what we mean by ``negligibly small'' and to reflect the maximal metric mismatch allowed by the RNC application in question.

Consider a metric-dependent observable, for instance, the length of a curve, which is to be given up to a precision $\tilde \varepsilon$. In a measurement process, this corresponds to an observer with a metric-responsive detector of resolution $\tilde \varepsilon$ measuring this observable, for example, a ruler measuring the curve length. Given an observable depending linearly on the metric, we have in the smallness condition (\ref{2)_general}) a linear dependence on $\tilde \varepsilon$ as well with $\varepsilon = \tilde \varepsilon$. For the example of the curve length, we have a dependence of the observable on $\sqrt{\tilde \varepsilon}$ and therefore $\varepsilon = \tilde \varepsilon^2$. Due to the resulting precision $\varepsilon$ in the metric, higher order terms in the metric expansion contributing less than $\varepsilon$ can be neglected in all calculations (their contribution is below the precision $\tilde \varepsilon$ of the observable and the resolution of the corresponding detector). Conversely, for a given adiabatic order, the precision $\tilde \varepsilon$ of the observable restricts the applicability of the RNC to some spacetime patch of finite size around the origin, since for greater distances terms of higher adiabatic order in the metric series will become too large and (\ref{2)_general}) will be violated for the given~$\varepsilon(\tilde \varepsilon)$.

The patch size determined by (\ref{2)_general}) does not always give the real domain of validity for $g^{(n)}$, however. In regions of quickly varying background curvature, such as for reference points near a physical singularity, terms of adiabatic order larger than $n+1$ in the metric expansion which depend on higher order derivatives of the Riemann tensor will already become important for small distances to the reference point. If $\varepsilon$ is chosen too large in such cases, the patch determined by (\ref{2)_general}) will extend far enough for the higher order terms, which are neglected in (\ref{2)_general}), to contribute considerably in the metric series. The real error introduced by neglecting terms of order higher than $n$ therefore grows much larger than $\varepsilon$ within the patch given by (\ref{2)_general}). When setting up RNC patches given such a quick varying of the curvature, we can show the insufficiency of (\ref{2)_general}) in this case by checking for physically unreasonable results after translating the RNC patch size to some other $x$ frame according to Step~5 (as we will show explicitly in Sec.~\ref{subsec: PG}). This reasoning can be difficult to employ, however, as there is no way to determine \textit{a priori} which physical quantities are suitable for this assessment. Nevertheless, if we can thus ascertain the necessity to consider higher orders of the metric expansion and calculating the patch size with (\ref{2)_general}) for some higher order $N>n$ is undesirable, for instance, because using $g^{(N)}$ leads to extensive computational effort in further calculations, we extend (\ref{2)_general}) and demand instead
\begin{eqnarray}
\label{2)_general_extended}
&& \left|\mathcal{O}_{g_{\alpha\beta}}(\xi^{n+k}) + \cdots + \mathcal{O}_{g_{\alpha\beta}}(\xi^{n+1})\right| \leq \varepsilon\, \left|{g_{\alpha\beta}}^{(n)}(\xi)\right|,
\end{eqnarray}
with suitable $\mathbb{N}\ni k\geq 2$ [for $n=0$ and $k=2$ this is equivalent to (\ref{2)_general})]. The left-hand side in (\ref{2)_general_extended}) equals $\left|{g_{\alpha\beta}}^{(n+k)}(\xi)-{g_{\alpha\beta}}^{(n)}(\xi)\right|$, and we would achieve the exact patch size for $k\rightarrow\infty$. Note that the triangle inequality should not be employed here, because $\mathcal{O}_{g_{\alpha\beta}}(\xi^{n+i})$ and $\mathcal{O}_{g_{\alpha\beta}}(\xi^{n+j})$ for $i \not = j$ can enter with different signs, and therefore the resulting patch size would be smaller than it actually is.

If spacetime regions of very quickly varying curvature are to be covered by the RNC, large $k$ [or, if (\ref{2)_general}) is to be used, much larger $n$ than initially desired] may be necessary to achieve the patch size. Calculating the metric series up to much higher orders than initially desired may be inconvenient, however. In this case, we can instead also reduce the maximal error $\varepsilon$ we allow for the approximation, thus ensuring that the patch will not reach too far and that higher order terms are negligible. To obtain this upper bound on $\varepsilon$ for some $k'$ smaller than what would actually be required for (\ref{2)_general_extended}) to give the patch size, we proceed as follows: First, we calculate the patch sizes using (\ref{2)_general_extended}) with $k=k'$ and $k=1$ [note that $k=1$ corresponds to employing (\ref{2)_general})]. We will denote the RNC configurations which satisfy the above conditions and therefore lie within the patches by $\xi^{(n)}_{\backslash (n+k')}$ and $\xi^{(n)}_{\backslash (n+1)} = \xi^{(n)}$, respectively. In the subscript, we denote the terms of maximal order $n+k'$ (and $n+1$) taken into consideration by ``dropping them'' from the metric series by $\backslash (n+k')$ [and $\backslash (n+1)$]. Second, we compare $\xi^{(n)}_{\backslash (n+k')}$ with $\xi^{(n)}$ and demand they agree up to $\mathcal{O}(\varepsilon)$. This corresponds to the patch obtained using (\ref{2)_general}) being small enough such that terms until order $n+k'$ remain negligible. Third, if we understand $\xi^{(n)}_{\backslash (n+k')}$ and $\xi^{(n)}$ as functions of $\varepsilon$, we can obtain the upper bound on $\varepsilon$ for order $n$ with terms only up to order $n+k'$ taken into account by requiring the aforementioned accordance of the patch sizes: $\left(\xi^{(n)}_{\backslash (k)} / \xi^{(n)}\right)\!(\varepsilon) = 1 + \delta$, demanding $|\delta|\ll \varepsilon$.

When comparing the order $n+1$ terms $\mathcal{O}_{g_{\alpha\beta}}(\xi^{n+1})$ with $g^{(n)}_{\alpha\beta}(\xi)$, we want to take into account that the metric tensor itself is not an observable. Rather, the metric is completely contracted in the action of physical systems such as point particles or scalar fields. This action then serves as the starting point of calculations which yield metric-sensitive observables. Taking this into account allows us to deal with certain pathological behaviors of the truncated metric expansion. For example, it is, in fact, possible that an off-diagonal component $g^{(n)}_{\alpha\beta}(\xi)$ becomes small compared to $\mathcal{O}_{g_{\alpha\beta}}(\xi^{n+1})$ in some RNC regions or that $g^{(n)}_{\alpha\beta}(\xi)$ even vanishes for certain $\xi$ configurations. We see that in such cases the right-hand sides of (\ref{2)_general}) and (\ref{2)_general_extended}) vanish, resulting in minimal or even vanishing $\xi$ on the left-hand side. In the action, such minimal contributions (at order $n$) will be irrelevant, however, and the minimal patch sizes derived thereof are consequently too restrictive. We therefore compare $\mathcal{O}_{g_{\alpha\beta}}(\xi^{n+1})$ with the maximum of all components at order $n$ which have to occur in the action as well, i.e., the corresponding diagonal terms. Thus, Eq.~(\ref{2)_general}) becomes
\begin{equation}
\label{2)_general_summation_rule}
\left|\mathcal{O}_{g_{\alpha\beta}}(\xi^{n+1})\right| \leq \varepsilon\,\text{max}_{\alpha,\beta}^\text{diag}\left\lbrace \left|g^{(n)}_{\alpha\beta}(\xi)\right|\right\rbrace,
\end{equation}
where we use max$_{\alpha,\beta}^\text{diag} \left\lbrace g_{\alpha\beta}\right\rbrace$ as a shortened notation for max$\left\lbrace g_{\alpha\beta},\, g_{\alpha\alpha},\, g_{\beta\beta}\right\rbrace$ with $g_{\alpha\alpha}$ and $g_{\beta\beta}$ denoting the diagonal components to an off-diagonal component $g_{\alpha\beta}$ $(\alpha\neq\beta$). {Diagonal components therefore still get compared only with each other.}

Condition (\ref{2)_general_summation_rule}) can, of course, also be extended analogously to (\ref{2)_general_extended}), and we have
\begin{eqnarray}
\label{2)_general_extended_summation_rule}
&& \left|\mathcal{O}_{g_{\alpha\beta}}(\xi^{n+k}) + \mathcal{O}_{g_{\alpha\beta}}(\xi^{n+k-1}) + \cdots + \mathcal{O}_{g_{\alpha\beta}}(\xi^{n+1})\right| \nonumber \\
&& \leq \varepsilon\,\text{max}_{\alpha,\beta}^\text{diag}\left\lbrace \left|g^{(n)}_{\alpha\beta}(\xi)\right|\right\rbrace,
\end{eqnarray}
with again some appropriate $\mathbb{N}\ni k\geq 2$.

Other metric contractions than the action will yield different comparison methods. If the specific contraction of the metric is unclear or too complicated, one is confined to comparing $\mathcal{O}_{g_{\alpha\beta}}(\xi^{n+1})$ just with $g^{(n)}_{\alpha\beta}(\xi)$ as given in (\ref{2)_general}) [or (\ref{2)_general_extended})]. If in this case $g^{(n)}_{\alpha\beta}(\xi)$ vanishes, one compares $\mathcal{O}_{g_{\alpha\beta}}(\xi^{n+1})$ [or the corresponding sum in (\ref{2)_general_extended})] with the minimum of the truncated sums of all other components.

After Step~2 it is already possible to employ the RNC within the determined domain of validity. One only has to take into account that using a truncated metric to calculate other tensors $T(g^{(n)}(\xi))$ will result in expansions of the $T$ that are not only truncated as well, but also contain additional terms if $T$ does not depend linearly on $g$. These terms can never match the usual Taylor expansions of the $T$. We will show this in detail for the Riemann tensor in Sec.~\ref{subsec: RNC_applied}. One therefore has to compute every tensor of interest individually and use the resulting expression instead of a Taylor expansion. If one does not wish to take the mismatch terms into account, Steps~3 and 4 are required.\\

\noindent\textbf{\textit{Step 4}}\\
\indent After having computed all other tensors $T(g^{(n)}(\xi))$ of interest in Step~3, we demand the obtained expressions coincide with the usual Taylor expansions. Since $T(g^{(n)}(\xi))$ was calculated using a metric which was truncated after order $n$, any $T(g^{(n)}(\xi))$ can only agree with its usual Taylor series up to some order $m\leq n$. We therefore require that the order $m+1$ terms, both of the usual Taylor expansion and of $T(g^{(n)}(\xi))$, be small compared to the Taylor expansion of $T$ truncated after order~$m$. The higher order terms of $T(g^{(n)}(\xi))$ are the aforementioned mismatch terms. This demand for the Taylor series and $T(g^{(n)}(\xi))$ is analogous to (\ref{2)_general}) for the metric.

Here, the problem of small or vanishing Taylor expansions of some $T$ components up to order $m$ and corresponding nonvanishing order $m+1$ terms of the Taylor expansion and/or $T(g^{(n)}(\xi))$ may also occur. In theory, one could again develop for all $T$ a comparison method that is analogous to the one for the metric and depends on the possible truncations of $T$ in the specific application. However, there is no distinguished contraction of an arbitrary $T$, as the action for the metric, and such summations of the $T$ components can become arbitrarily complex. As a consequence, we will here, whenever some Taylor expansion up to order $m$ vanishes, compare the corresponding nonvanishing order $m+1$ terms with the minimum of all nonvanishing Taylor expansions.

All in all, we find two sets of restrictions for every $T$, one from the Taylor expansion and one from $T(g^{(n)})$. The patch size is then determined as the minimum of the conditions from Steps~2 and 4. The reason behind this is that we used the conditions from Step~2 to restrict the RNC patch such that the metric mismatch is small. Thus we ensured the validity of the truncated metric used to calculate $T(g^{(n)})$ in Step~3. The additional conditions from Step~4 therefore implicitly require those from Step~2. It is furthermore important to note that, as a consequence, the conditions from Step~4 together with Step~2 will always be more restrictive than those from Step~2 alone. This is because using an already truncated metric expansion to calculate $T(g^{(n)})$ introduces yet another mismatch compared to the full expansion of $T$.\\

\noindent\textbf{\textit{Step 5}}\\
\indent Simply plugging the RNC conditions into the coordinate transformation (\ref{RNC_coord_trafo}) will produce unreasonable results, as the RNC observer is in general located at some different reference point than the observer corresponding to the $x$ coordinates. For the same reason, one can in general not calculate the extent of the RNC patch along some $\gamma_v$ using its proper length $\tau$ as a curve parameter. $\tau$ only serves as a clock for an observer moving along $\gamma_v$. Any other observer comes with a different clock (eigentime) that they use as curve parameters. 

In order to translate the patch size to another observer in $x$ coordinates with the eigentime $x^0 = t_{\text{obs}}$, i.e, to examine the $x$ coordinate patch in which the other observer can describe physics using RNC, we therefore proceed as follows: First, we choose some geodesic $\gamma_v$ employed for the exponential map (in $x$ coordinates) along which we wish to determine the patch size. As discussed in Sec.~\ref{sec: prelim}, $\gamma_v$ gets mapped onto a straight line in RNC that is given by $\xi^\alpha = \lambda^\alpha\tau$, where we find the transformed 4-velocity $\lambda^\alpha = v^a e^\alpha_a$ using the inverse vierbein.

Second, we determine the maximal proper length along $\gamma_v$ within the RNC patch $\tau^{(n)}_{\gamma_v}$ by requiring the corresponding line in the RNC to remain in the domain of validity. For that, we consider the neighborhood $\Sigma^{(n)}_\varepsilon$ of $\xi^{(n)}$ around the origin in RNC allowed by the $\xi$~conditions computed in the previous steps. This patch in normal coordinate space is the domain of validity. The RNC patch's boundary is described as an implicit surface by $\partial\Sigma^{(n)}_\varepsilon(\xi^\alpha)=0$. It marks the RNC~region corresponding to the maximal error $\varepsilon$ one wants to allow for the approximation and therefore gives the maximal $\xi^{\alpha\,(n)}$. We now obtain $\tau^{(n)}_{\gamma_v}$ by computing the intersection of the straight line corresponding to $\gamma_v$ with this surface $\partial\Sigma^{(n)}_\varepsilon\!\!\left(\lambda^\alpha \tau^{(n)}_{\gamma_v}\right) =0$.

Third, we reparametrize $\gamma_v(\tau_{\gamma_v})$ by the observer's eigentime $t_{\text{obs}}$ and compute the maximal eigentime $t^{(n)}_{\text{obs}}(\tau^{(n)}_{\gamma_v})$ along this geodesic. By plugging this into $\gamma_v(t_{\text{obs}})$ we obtain the maximal extent of the RNC patch along this geodesic $\gamma_v$ as seen by the given observer. 

We will elaborate the dependence of patch sizes on the observer in detail for the geometry of a Schwarzschild black hole in Sec.~\ref{sec: BlackHole}.

\section{Riemann Normal Coordinates}
\label{sec: RNC}
\subsection{Applying the method for order $n=2$}
\label{subsec: RNC_applied}

We now provide sample calculations for Steps~1 to~4 for an expansion of the metric up to $n=2$ and the Riemann tensor being the tensor of interest. Also, we employ condition (\ref{2)_general_summation_rule}) for the metric. We thus derive the restrictions defining the patch in which the metric, contracted in an action or a comparable object, and the curvature tensor are approximated well by the RNC metric expanded up to order 2. In Sec.~\ref{sec: BlackHole} we will apply these conditions to the geometry of a Schwarzschild black hole and perform Step~5 in detail.\\

\noindent\textbf{\textit{Step 1}}\\
\indent From (\ref{geom_id1}) with $m=0$, one obtains the first order coefficient of the metric expansion $g_{\alpha\beta,\mu} = \Gamma_{(\alpha\beta)\mu} =0$. For the second order coefficient one uses $g_{\alpha\beta,\mu\nu} = \Gamma_{(\alpha\beta)\mu,\nu} =0$ and plugs (\ref{geom_id2}) into (\ref{geom_id1}) with $m=1$. An analogous calculation using (\ref{geom_id3}) and (\ref{geom_id1}) with $m=2$ yields the third order term of the metric (and correspondingly for higher orders). The metric up to order 3 $g^{(3)}(\xi)$ is given in (\ref{RNC_met_intro}).\\
	
\noindent\textbf{\textit{Step 2}}\\
\indent For (\ref{RNC_met_intro}) up to order 2 to be the correct metric to be used in an action, we demand that (\ref{2)_general_summation_rule}) hold. This gives the condition
\begin{equation}
\label{2)_general_order2}
\left| R_{\alpha\mu\beta\nu,\varrho} \xi^\mu \xi^\nu \xi^\varrho \right| \leq 2 \varepsilon\,\text{max}_{\alpha,\beta}^\text{diag}\left\lbrace
\left|3 \eta_{\alpha\beta} -  R_{\alpha\mu\beta\nu}\xi^\mu \xi^\nu \right|\right\rbrace ,
\end{equation}
with the notation $\text{max}_{\alpha,\beta}^\text{diag}$ as introduced in (\ref{2)_general_summation_rule}).

In general, the easiest method to deduce concrete\linebreak $\xi$~restrictions from conditions such as (\ref{2)_general_order2}), is to first consider the conditions along the axes, i.e., to determine $\left.\xi^{\alpha\,(n)}\right|_{\xi^\mu= \,0}$, $\forall\,\mu\neq\alpha$. Then, an iterative procedure considering all possible $\xi^\alpha$-$\xi^\beta$~combinations with only two, one, and finally no $\xi$ set to 0 that only alters the axial conditions when necessary produces the final $\xi$~conditions defining $\Sigma^{(n)}_\varepsilon$. We will show this in detail in Sec.~\ref{subsec: PG}.\\

\noindent\textbf{\textit{Step 3}}\\
\indent We compute the Riemann tensor $R_{\alpha\beta\gamma\delta}(g^{(2)}(\xi))$ as the tensor of interest. Using the intermediate results $\Gamma_{\alpha\beta\gamma}(g^{(2)}(\xi)) = -1/3\, R_{\alpha(\beta\gamma)\mu}\xi^\mu$ and $\Gamma^\alpha{}_{\beta\gamma}(g^{(2)}(\xi)) = -1/3\, R^\alpha{}_{(\beta\gamma)\mu}\xi^\mu - 1/9\, R^\alpha{}_{\mu\sigma\nu} R^\sigma{}_{(\beta\gamma)\varrho}\xi^\mu\xi^\nu\xi^\varrho$ we obtain
\begin{eqnarray}
\label{Riem_order2}
&& R_{\alpha\beta\gamma\delta}(g^{(2)}(\xi)) = R_{\alpha\beta\gamma\delta}  \nonumber \\
&& + \frac{1}{9}(R_{\alpha(\gamma\varrho)\mu}R^\varrho{}_{(\beta\delta)\nu} - \gamma \leftrightarrow \delta)\xi^\mu\xi^\nu + \cdots\;.
\end{eqnarray}
The neglected terms contain the contraction of two Riemann tensors with two $\xi$'s in higher orders.\\

\noindent\textbf{\textit{Step 4}}\\
\indent Comparing (\ref{Riem_order2}) with the usual Taylor expansion $R_{\alpha\beta\gamma\delta}(\xi) = R_{\alpha\beta\gamma\delta} + R_{\alpha\beta\gamma\delta,\mu}\xi^\mu + \cdots$ and demanding they coincide, we find the two additional conditions
\begin{eqnarray}
\label{4)_general_Riem_1}
&& \left|(R_{\alpha(\gamma\varrho)\mu}R^\varrho{}_{(\beta\delta)\nu} - \gamma \leftrightarrow \delta)\xi^\mu\xi^\nu\right| \leq 9\varepsilon \left|R_{\alpha\beta\gamma\delta}\right|, \\
\label{4)_general_Riem_2}
&& \left|R_{\alpha\beta\gamma\delta,\mu}\xi^\mu\right| \leq \varepsilon\left| R_{\alpha\beta\gamma\delta}\right|.
\end{eqnarray}

We used here the same $\varepsilon$ as for the metric in (\ref{2)_general}), but since the precision requirement for the Riemann tensor may correspond to the resolution of a different observable, we can in principle also have another maximal error. It is here important to note that (\ref{4)_general_Riem_1}) compares terms of quadratic order in $\xi$ with $\xi$ independent terms, meaning it will restrict the domain of validity only with a factor $\sqrt{\varepsilon}$, while in (\ref{4)_general_Riem_2}) a factor $\varepsilon$ enters. Also, from the calculations leading to (\ref{Riem_order2}) we see that using only the Minkowski metric $\eta=g^{(0)}$ yields a nonexistent patch size for the Riemann tensor, because the metric derivatives vanish in this case. This shows that the domain of validity for the coordinate independent Kretschmann scalar is also nonexistent for $n=0$, whereas for $n=2$ the correct value is reestablished at the reference point. This reflects the fact that obtaining curvature information of a manifold always requires a neighborhood of finite size.

The patch size from Step 4 together with Step 2 is now determined by the minimum of the $\xi$ over the conditions (\ref{2)_general_order2}), (\ref{4)_general_Riem_1}), and (\ref{4)_general_Riem_2}).

\subsection{Literature estimate for the patch size}
\label{subsec: RNC_estimate}

We can establish the connection between our concrete calculations for the domain of validity and the patch size estimate from the literature by means of several trivializing estimates for the $\eta$~patch size determined by (\ref{2)_general_extended}) with $n=0$ and $k=3$. Let us first demand that not the sum of $\mathcal{O}_{g_{\alpha\beta}}(\xi^2)$ and $\mathcal{O}_{g_{\alpha\beta}}(\xi^3)$ be negligible compared to the Minkowski metric, but rather that this holds for both terms individually. Also, let us consider only the dependence of the terms on the Riemann tensor and its derivatives as well as on powers of $\xi$, i.e., we will neglect all prefactors and all index contractions. Furthermore, since $\eta_{\alpha\beta}\lambda^\alpha\lambda^\beta = 1$ restricts the components $\lambda^\alpha$ to be maximal $\pm 1$, let us set $\lambda$ to $1$ for each direction in order to consider the maximal extent of the $\eta$~patch independent of directions. As a consequence, we can consider $\tau^{(0)}$ instead of $\xi^{\alpha\,(0)}$. In order to avoid overestimating the patch size, let us compare all nonvanishing curvature components and derivatives with each other. Finally, let us also ignore the dependence on $\varepsilon$ and only demand general smallness of higher order terms. We thus find the estimate for the RNC patch size from the literature (see, for example, \cite{Domain2, Nesterov:1999ix}), given as a restriction on the distance from the reference point along any geodesic:
\begin{equation}
\label{patch_size_lit}
\tau^{(0)} \ll \text{min}\left\lbrace \frac{1}{\sqrt{|R_{\alpha\beta\gamma\delta}|}},\, \frac{|R_{\alpha\beta\gamma\delta}|}{|R_{\mu\nu\varrho\sigma,\chi}|} \right\rbrace\,\,\, \forall\, \alpha,\,\beta,\,\ldots,\,\chi\,.
\end{equation}
The minimum is here to be taken over all possible nonvanishing components of the Riemann tensor and its derivatives, i.e., over all possible combinations of the independent indices $\alpha,\,\ldots,\,\chi$ which describe nonzero components.

In the literature, these two conditions on the metric are found by estimating the RNC validity using the curvature radius, i.e., the length scale at which, for example, geodesic deviation becomes important, and demanding curvature to not change significantly within the patch compared to the reference point. Since higher order parameters of the metric expansion are given by higher derivatives of the Riemann tensor and polynomials of lower order parameters, it is then argued that (\ref{patch_size_lit}) ensures that $\eta$ will always be the dominant term of the metric expansion.

We see, however, that greatly simplifying assumptions were necessary to reach the patch size estimate (\ref{patch_size_lit}) from our concrete conditions and that the usual reasoning behind this estimate is also based on uncertain assumptions. Furthermore, (\ref{patch_size_lit}) only gives a rough estimate of the domain of validity instead of a concrete maximal value for $\tau^{(0)}$ and only describes the $\eta$~patch size. It cannot account for metric expansions up to higher orders, which will be valid on larger patches.

In Sec.~\ref{subsec: PG} we will compare the results of (\ref{patch_size_lit}) with our results obtained using (\ref{2)_general_extended_summation_rule}) with $n=0$ and $k=3$ for the geometry of a Schwarzschild black hole.

\section{Fermi Normal Coordinates}
\label{sec: FNC}

Fermi normal coordinates as developed in \cite{Manasse:1963zz} are another set of normal coordinates. Their construction relies on a given geodesic $\gamma(\tau)$. If the geodesic $\gamma$ is obtained along some interval by solving the geodesic equation in $x$~space given some initial conditions, the usual FNC prescription along this geodesic can be employed.

\begin{figure}
\includegraphics{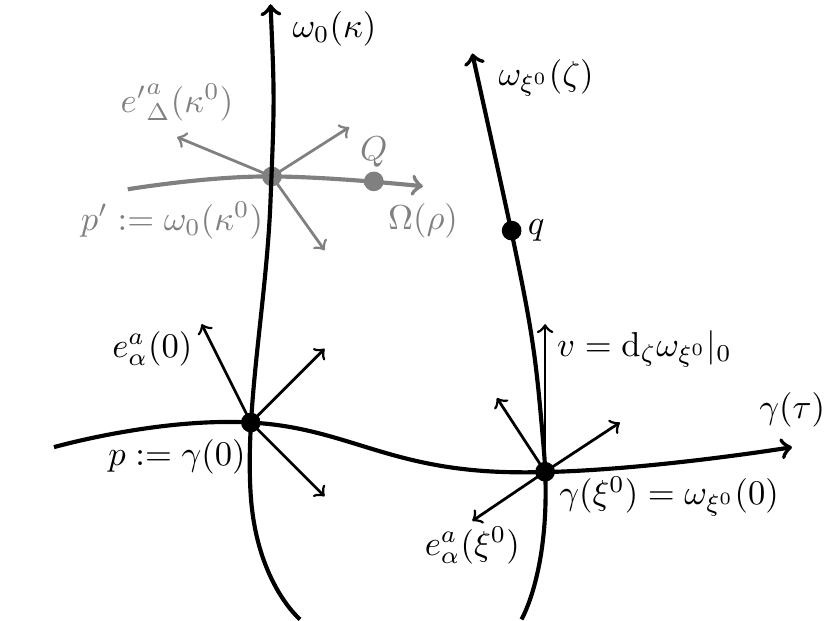}%
\caption{\label{fnc} The dark part depicts the FNC construction along the geodesic $\gamma(\tau)$ with orthogonal RNC expansions at the points $p$ and $\gamma(\xi^0)$ with vierbein $e_\alpha^a(0)$ and $e_\alpha^a(\xi^0)$. An additional RNC expansion at $p'$ with vierbein ${e'}^a_\Delta(\kappa^0)$ is depicted in lighter gray to illustrate the alternative patch size calculation for the ``FNC around a point'' construction as discussed in Appendix~\ref{sec: appendix a}.}
\end{figure}

\subsection{Fermi normal coordinates along a geodesic}
\label{subsec: FNC_geodesic}

Given some arbitrary reference point on the geodesic $p = \gamma(0)$, the FNC assigned to a point $q$ are obtained as illustrated in Fig.~\ref{fnc}: Starting at $p$, $\gamma$ is followed for the length $\tau=\xi^0$ until $\gamma(\xi^0)$, where an RNC expansion is performed in the orthogonal directions. These orthogonal RNC are represented by the geodesic $\omega_{\xi^0}(\zeta)$ with $\omega_{\xi^0}(0) = \gamma(\xi^0)$ and $\left.\mathrm{d}_\zeta\omega_{\xi^0}\right|_0 = v\perp \left.\mathrm{d}_\tau\gamma\right|_{\xi^0}$ that reaches $q$ after some length $\zeta^0$. Since the RNC expansion is  orthogonal and describes the spatial part of the FNC, we will label the associated coordinates by indices $\bar\alpha$ and distinguish them from $\xi^0$ coming from the central geodesic $\gamma$. The point $q$ then receives the coordinates $(\xi^0,\xi^{\bar 1},\xi^{\bar 2},\xi^{\bar 3})$ with $\xi^{\bar\alpha} = \zeta \lambda^{\bar\alpha}$, where the $\lambda^{\bar\alpha}$ are obtained as for RNC by expanding $v$ in terms of the vierbein at $\gamma(\xi^0)$, $v^a = \lambda^{\bar\alpha} e^a_{\bar\alpha}(\xi^0)$, and therefore satisfy $\eta_{\bar\alpha\bar\beta}\lambda^{\bar\alpha}\lambda^{\bar\beta} = 1$.
	
The vierbein $e^a_\alpha(\xi^0)$ at $\gamma(\xi^0)$ is obtained from an initial vierbein $e^a_\alpha$ at $p$ by parallel transport. For this initial vierbein one chooses $e^a_0 = \left.\mathrm d_\tau\gamma\right|_0$ and fixes the remaining $e^a_{\bar\alpha}$ by requiring orthonormality. By this construction $e^a_0(\tau) = \mathrm d_\tau\gamma(\tau)$ holds on the whole of $\gamma$ and $v$ is expanded only in terms of the ``orthogonal part'' of the vierbein.

By this construction, the interval of the central geodesic $\gamma$ gets mapped onto the line $(\xi^0,0,0,0)$ with $\xi^0\in U\subseteq\mathbb{R}$ and the orthogonal geodesics get mapped onto the straight lines $\zeta(0,\lambda^{\bar 1},\lambda^{\bar 2},\lambda^{\bar 3})$ at every point $\gamma(\xi^0)$. Thus, FNC cover a tubular region around the central geodesic.
	
Due to the reliance of FNC on the RNC construction, their coordinate transformation and metric expansion are, in close analogy to (\ref{RNC_coord_trafo}) and (\ref{RNC_met_intro}), given by 
\begin{eqnarray}
\label{FNC_coord_trafo}
&& x^a = p^a + \gamma(\xi^0) + \xi^{\bar\alpha}e^a_{\bar\alpha}(\xi^0) + \cdots\;, \\
\label{met_FNC}
&& g_{\alpha\beta}(\xi) = \eta_{\alpha\beta} - G(\alpha,\beta)R_{\alpha\bar{\mu}\beta\bar{\nu}}(\xi^0)\xi^{\bar\mu}\xi^{\bar\nu} + \cdots\;,
\end{eqnarray}
with symmetric $G$ defined as $G(0,0) = 1$, $G(0,\bar\alpha) = 2/3$, and $G(\bar\alpha,\bar\beta) = 1/3$ (see \cite{Manasse:1963zz}). Since every point $\gamma(\xi^0)$ of the central geodesic serves as a reference point for an orthogonal RNC expansion, the Riemann tensor (and its derivatives appearing in higher order terms of the metric) have to be evaluated at every $\gamma(\xi^0)$. Therefore, additional geometrical information is required along the interval of $\gamma$ in contrast to a single reference point as for RNC.

Notice $G(\bar\alpha,\bar\beta)=1/3$, which encodes both FNC containing standard RNC expansions in the orthogonal directions at every $\xi^0$ and the tubelike ``shape of FNC.'' From this it also follows that we can use our previous results on RNC patch sizes to find the domain of validity of FNC. The extent of a tubelike region is naturally given by its diameter. In case of the FNC tube, this diameter is constituted by the time-dependent orthogonal validity around every $(\xi^0,0,0,0)$, which is in turn given by the time-dependent patch sizes of the orthogonal RNC patches for every $\xi^0$.

For the explicit calculations we therefore restrict ourselves to the spatial part $g_{\bar\alpha\bar\beta}$ of the metric (\ref{met_FNC}) and employ it for Steps 1 to 4 from RNC. This yields conditions analogous to (\ref{2)_general_order2}), (\ref{4)_general_Riem_1}), and (\ref{4)_general_Riem_2}) for every $\xi^0$; i.e., they are given in terms of $R_{\bar\alpha\bar\beta\bar\gamma\bar\delta}(\xi^0)$ instead of $\left.R_{\alpha\beta\gamma\delta}\right|_p$. Since FNC follow the central geodesic for some time interval or even in its entirety, integrating the $\xi^0$ dependent patch sizes over this time results in the tubelike domain of validity for FNC.

In other coordinate systems the time-dependent RNC patch will be more intricately shaped, but the concept of integrating over the time-dependent patch size of an RNC patch which follows the central geodesic $\gamma$ to obtain the cylindrical domain of validity of FNC remains.
	
\subsection{Fermi normal coordinates around a point}
\label{subsec: FNCP}

For the FNC as discussed above, an interval of the central geodesic $\gamma$ is required. Solving the geodesic equation for arbitrary spacetimes or arbitrary initial conditions can in general be infeasible, however. In this case and in order to still take the geodesic into account as a preferred geodesic of the physical system, we can temporally Taylor expand $\gamma$ in $\tau=\xi^0$ around $p$ and then employ again the familiar FNC prescription. We will here assume the order of the $\xi^0$~expansion and the resulting temporal validity, together with the corresponding error, to be determined independently and to enter the FNCP construction as external parameters. The vierbein and Riemann components at $\gamma(\xi^0)$, which were before obtained by parallel transport along $\gamma$, are now found by parallel transport along the Taylor expanded $\gamma$; i.e., we Taylor expand them in $\xi^0$ as well. Using (\ref{FNC_coord_trafo}) and (\ref{met_FNC}) and denoting the $\tau$~differentiation by a dot, we then obtain the coordinate transformation and metric expansion of the resulting temporally Taylor expanded FNC
\begin{eqnarray}
\label{FNC_coord_trafo_alt}
\!\!\!\!\!\!\!&& x^a(\xi) = p^a + e^a_0 \xi^0 + \dot e^0_a(\xi^0)^2 + \xi^{\bar\alpha}(e^a_{\bar\alpha} + {\dot e}^a_{\bar{\alpha}}\xi^0) + \cdots\, ,\, \\
\label{met_FNC_alt}
\!\!\!\!\!\!\!&& g_{\alpha\beta}(\xi) = \eta_{\alpha\beta} - G(\alpha,\beta)(R_{\alpha\bar\mu\beta\bar\nu} \nonumber \\
\!\!\!\!\!\!\!&& \phantom{g_{\alpha\beta}(\xi) =}\, + {\dot R}_{\alpha\bar\mu\beta\bar\nu}\xi^0)\xi^{\bar\mu}\xi^{\bar\nu} + \cdots \;.
\end{eqnarray}
We call this procedure an ``FNC expansion around a point.''
	
In order to determine the domain of validity of these FNCP around $p$, we proceed analogously to the usual FNC along a geodesic. This means that given an expansion up to some order in $\xi^0$ and the corresponding temporal validity, we can proceed analogously to regular FNC and use the spatial part of (\ref{met_FNC_alt}) to calculate the spatial extent along the approximated central geodesic. This will again yield conditions (\ref{2)_general_order2}), (\ref{4)_general_Riem_1}), and (\ref{4)_general_Riem_2}) depending now on $R_{\bar\alpha\bar\beta\bar\gamma\bar\delta} + {\dot R}_{\bar\alpha\bar\beta\bar\gamma\bar\delta}\xi^0 + \cdots$ instead of the full $R_{\bar\alpha\bar\beta\bar\gamma\bar\delta}(\xi^0)$. The domain of validity of FNCP is therefore a temporally restricted part of the full tubular region covered by FNC along $\gamma$. Since cutting the expansion in $\xi^0$ introduces yet another mismatch to the full series, this procedure is accompanied by another error. Thus, fixing the total error, assembled by the temporal expansion and the RNC expansion in the orthogonal direction, leads to a tubular region which shrinks in the orthogonal direction for later times to a point at the maximal possible time. The same behavior occurs for negative times and thus the patch validity can be described by a point in space growing in time to a finite ball shaped region and eventually shrinking again to a point.

If no \textit{a priori} knowledge of the temporal validity is available, the $\xi^0$~expansion has to be treated in the same way as the orthogonal $\xi^{\bar\alpha}$~expansions. Therefore, FNCP become comparable to RNC in the sense that they are then simply another way of assigning normal coordinates to a spacetime patch while using only geometrical information at a single reference point. The only difference is the remaining preferred direction along $\gamma$ of the FNCP. As fits intuition, we then find the validity of these FNCP to be that of a corresponding RNC patch around said reference point. The detailed calculations showing this are quite lengthy, however, and we therefore postpone this discussion to Appendix~\ref{sec: appendix a}.

It should be noted that FNC can also be constructed in the presence of nongravitational forces, as was recently shown in \cite{Frauendiener:2018gkw}.\footnote{We wish to thank the reviewer for pointing out this article and the connection to our work.} Due to the resulting acceleration, the observer's worldline is no longer a geodesic and the parallel transport of the vierbein along the wordline has to be substituted by the Fermi-Walker transport. For the sake of simplicity, however, we avoid nongravitational effects throughout this article. Furthermore, in \cite{Frauendiener:2018gkw} one also finds restricting assumptions on the observer's eigentime that essentially correspond to employing FNCP at leading order of the $\xi^0$~expansion.

\section{Patch Sizes in the Geometry of a Schwarzschild Black Hole}
\label{sec: BlackHole}

Having extensively discussed our method for determining the domain of validity of different normal coordinate systems, we will now exemplarily employ it to determine RNC patch sizes in the geometry of a Schwarzschild black hole. We choose this geometry for its prominence and also because the occurrence of a singularity and of an event horizon within the geometry allow for the thorough verification of our calculated patch sizes.

We will describe the geometry of the Schwarzschild black hole by Schwarzschild and Painlev\'e-Gullstrand \mbox{$x$ coordinates} and discuss RNC patches obtained from the metric and the Riemann tensor. In detail, we will present explicit conditions and patch size plots for the $\eta$~patch -- obtained by using both (\ref{2)_general_summation_rule}) and (\ref{2)_general_extended_summation_rule}) with $k=3$ -- as well as for the $g^{(2)}$ patch -- here found by employing (\ref{2)_general_order2}). Additionally, we will give the patch size conditions for the Riemann tensor derived from (\ref{4)_general_Riem_1}) and (\ref{4)_general_Riem_2}) and we will see that they are indeed more restrictive than the conditions for the metric at order $n=2$. Finally, we will show the growth of the RNC patch by also plotting patch sizes for the $g^{(3)}$ and $g^{(4)}$ patches found using (\ref{2)_general_summation_rule}).

\subsection{Patch size observer dependence}
\label{subsec: observer_dep}

The Schwarzschild geometry can be addressed in various coordinate systems with different properties. Apart from minor differences such as between Cartesian and polar coordinates, each specific choice of the coordinate system corresponds to a unique observer. Therefore, if we restrict ourselves to timelike observers, the coordinate time equals the observer's eigentime. These different eigentimes result in crucially different structures of the causal future and past of the normal coordinate reference point $p$ as seen by different observers. For instance, the most common observer corresponding to the Schwarzschild coordinates can never observe a physical object crossing the event horizon from the outside, while for other observers this is in principle possible. Different observers will therefore describe the same physical objects with essentially different observations. This is also represented in the size of normal coordinate patches after they are translated to other observers.

The common line element for a Schwarzschild black hole with mass $M$ in spherical Schwarzschild coordinates $\{ x^t=t, x^r=r, x^\theta=\theta, x^\phi=\phi \}$ reads
\begin{equation}
\label{line_elem_SS}
\mathrm d s^2 = - f(r)\, \mathrm d t^2 + \frac{1}{f(r)}\,\mathrm d r^2 + r^2  \mathrm d \Omega^2,
\end{equation}
with $f(r) = 1 - 2M/r$ and $\mathrm d \Omega^2 = \mathrm d \theta^2 + \sin^2\theta\, \mathrm d \phi^2$. Here one has a coordinate singularity at the horizon $r=2M$. 

Due to this coordinate singularity, it is, as mentioned above, impossible to cross the horizon in these coordinates. As a probe we choose a radially, freely infalling object starting at rest at infinity. We can deduce the 4-velocity $v$ of such an object from the line element
\begin{equation}
\label{vel_SS_freefall}
v^a = \left( \frac{1}{f(r)}, - \sqrt{\frac{2M}{r}}, 0, 0 \right).
\end{equation}
Due to $v^t \stackrel{r\,\rightarrow\,\infty}{\rightarrow} 1$ in (\ref{vel_SS_freefall}), these coordinates correspond to an observer who is located at spatial infinity and whose coordinate eigentime equals the coordinate time $t_{\text{obs}}=~t$. This observer is called the Schwarzschild observer. Since the coordinate velocity of the probe approaches zero at the horizon $\mathrm d r/\mathrm d t  = v^r/v^t  \propto f(r) \stackrel{r\,\rightarrow\, 2M}{\rightarrow} 0$, the Schwarzschild observer measuring with their eigentime~$t$ does not see the probe crossing the event horizon in these coordinates.

For reasons of clarity, we restrict ourselves to global coordinate systems which are related by coordinate transformations solely in time. Therefore, the 4-velocity of the probe in some new coordinates is changed to have an arbitrary $v^t$ compared to (\ref{vel_SS_freefall}). Since at the event horizon $v^r$ is finite, $v^t$ has to diverge in order to prevent the probe from falling into the black hole, which is the case for the Schwarzschild observer. Thus, changing $v^t$ such that it remains finite outside the black hole ($r\geq 2M$) results in a coordinate time for which objects can cross the event horizon.

An interesting example is $v^t_{\text{PG}} = 1$ corresponding to the Painlev\'e-Gullstrand (PG) observer who follows the same geodesic as the freely infalling probe \cite{Martel:2000rn}. As a consequence, $t_{\text{obs}}=t_{\text{PG}}$ equals the eigentime of the probe and the PG observer's proper time differential is given by $\mathrm d t_{\text{PG}}^2  = f(r)\mathrm d t^2 - f^{-1}(r)\mathrm d r^2$, which equals the negative line element of the probe's geodesic $- \mathrm d s^2$. \pagebreak Dividing by~$\mathrm d s$ and using $v^a = \mathrm{d}x^a/\mathrm{d}s$, the proper time is then given by
\begin{equation}
\label{proper_time}
\mathrm d t_{\text{PG}} =  - v_t \mathrm d t - v_r \mathrm d r = \mathrm d t + \frac{\sqrt{\frac{2M}{r}}}{f(r)} \mathrm d r\;.
\end{equation}
Inserting this back into (\ref{line_elem_SS}), the line element in PG coordinates $\{x^{t_{\mathrm{PG}}} = t_{\mathrm{PG}},x^r = r,x^\theta = \theta ,x^\phi = \phi\}$ is found:
\begin{equation}
\label{line_elem_PG}
\mathrm d s^2_{\text{PG}} = - f(r)\, \mathrm d t_{\text{PG}}^2 + 2  \sqrt{\frac{2M}{r}} \,\mathrm d r \, \mathrm dt_{\text{PG}} + \mathrm d r^2 +  r^2\mathrm d \Omega^2\;.
\end{equation}
Due to the mixing between the temporal and the spatial parts in (\ref{proper_time}), the resulting line element and metric become nondiagonal. At the same time, however, the coordinate singularity of Schwarzschild coordinates is removed in PG coordinates. The normalized 4-velocity of the freely infalling, timelike probe and the PG observer is now given by
\begin{equation}
\label{vel_PG_freefall}
v^a_{\text{PG}} = \left( 1, - \sqrt{\frac{2M}{r}}, 0, 0 \right).
\end{equation}

We can now use either (\ref{vel_SS_freefall}) or (\ref{vel_PG_freefall}) to construct a vierbein and find the RNC system of the probe. In both cases we will find the same RNC, because the two initial coordinate systems describe the same geometry of a Schwarzschild black hole. Properties of specific coordinate systems such as coordinate singularities or the different structures of the causal future and past of $p$ as seen by the corresponding observers are not carried over to the RNC by construction. An RNC system depends only on the curvature information at the reference point and not on the properties of the initial observer such as the Schwarzschild and PG observer. As described in the preliminaries, any point within the RNC patch is uniquely addressed by a geodesic linking this point and the reference point. The eigentime of this geodesic is then used by the RNC observer to parametrize the distance between the point and the origin of the RNC. Thus, the dependence on the eigentimes of different observers is removed. For example, choosing the reference point to be outside a black hole and the point we wish to address to be inside, we can take an infalling geodesic and use its eigentime to find the distance between these points. This, however, corresponds to the scenario of taking PG coordinates, and therefore crossing the event horizon is achievable. The choice of this geodesic is independent of the observer's properties.

The very same behavior occurs in the FNC construction for the part orthogonal to the central geodesic. Since FNC rely on this geodesic in contrast to a single point, they depend on the properties of the geodesic and thus on the observer following it. As a consequence, only FNC of a geodesic which approaches the event horizon sufficiently closely or reaches into the black hole can access the interior of the black hole. We will discuss this and the properties of translated FNC patch sizes in Sec.~\ref{subsec: PG}.

Finally, since FNCP are a temporally Taylor expanded version of ordinary FNC, the observer dependence in the region of temporal validity is exactly the same.

Although the structure of the causal future and past of $p$ as seen by different observers corresponding to different initial coordinates is irrelevant as far as the patch size in normal coordinates is concerned, it is of crucial importance for the disparity of translated patch sizes in the initial $x$ coordinates as determined in Step~5. This is due to the fact that, as discussed above, coordinate transformations which include a change of observer and therefore also a change of eigentime potentially also change how this causal future and past are observed. Implementing the procedure given in Step~5, we will in the following investigate this for the examples of the Schwarzschild observer and the PG observer.

\subsection{Painlev\'e-Gullstrand observer}
\label{subsec: PG}

We begin by considering the PG observer and the associated PG coordinates. This observer's 4-velocity $v_{\text{PG}}$ is given by (\ref{vel_PG_freefall}). As discussed above, this observer can see objects crossing the horizon. The coordinate transformation from $\{t_{\text{PG}},r,\theta,\varphi\}$ to RNC $\{\xi^0,\xi^1,\xi^2,\xi^3\}$ is performed using some vierbein. Its components can, for example, be obtained by setting $e^a_0 = v^a_{\text{PG}}$ and fixing the other components according to the orthonormality condition $g^{\text{PG}}_{a b} e_\alpha^a e_\beta^b = \eta_{\alpha\beta}$. One then finds
\begin{equation}
\label{vierbein_PG}
e^a_0 = v^a_{\text{PG}}, \quad e^r_1 = 1, \quad e^\theta_2 = \frac{1}{r} , \quad e^\phi_3 = \frac{1}{r\sin\theta}\;.
\end{equation}

It is important to note that the RNC observer is characterized as freely falling, just as the PG observer for the black hole. Therefore, RNC are simply another coordinate system for the PG observer and RNC and PG coordinates share the same eigentime.

Performing a tensor transformation of the Riemann curvature tensor from PG coordinates to RNC, one finds the components $R_{\alpha\beta\gamma\delta} = R_{abcd} e^a_\alpha e^b_\beta e^c_\gamma e^d_\delta$ to be given by
\begin{eqnarray}
\label{Riem_RNC}
&& R_{2020} = R_{3030} = R_{1221} = R_{1331} = \frac{M}{r_0^3}\;, \nonumber \\
&& R_{0110} = R_{3232} = \frac{2M}{r_0^3}\;,
\end{eqnarray}
where $r_0$ is the radial value of the reference point $p$.

\begin{figure}
	\includegraphics[scale=1]{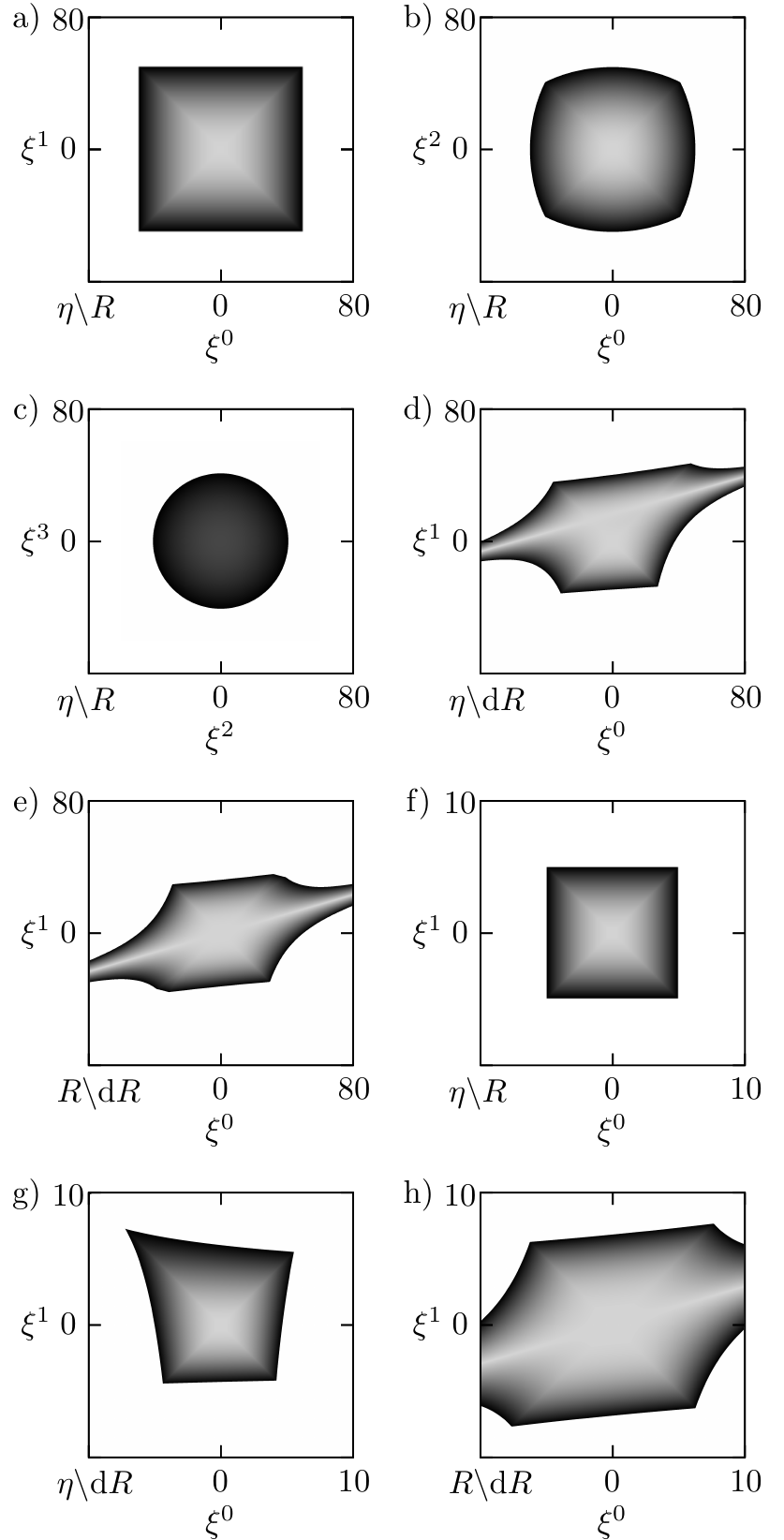}%
	\caption{\label{RNCpatch} 
		Minkowski patches resulting from neglecting the second order Riemann term are denoted by $\eta/R$ and depicted in a) - c) and f). Minkowski patches which are obtained by discarding the second and third order terms are denoted with  $\eta/\mathrm dR$ and given in d) and g). RNC patches of second order with the third order term discarded are denoted with $R/\mathrm dR$ and depicted in e) and h). Darker shades of gray correspond to increasing errors with the maximal error $\varepsilon$ reached in the black regions. White areas mark errors larger than $\varepsilon$ outside the domain of validity. The parameters for the plots a) - e) are $M=1$, $r_0=24$, and $\varepsilon=0.1$, whereas for f) - h) they differ with $\varepsilon=10^{-3}$. All directions $\xi^\mu$ not plotted are set to $0$ except in c), where $\xi^0$ is set to its maximal value of approximately $37$. The dependence of the patch size and shape on $\varepsilon$ and the used expansion terms is illustrated in a) and d)~-~h) in the $\xi^0$-$\xi^1$~plane.  These are characteristic patches leading to the patch size conditions for $\eta/R$, $\eta/\mathrm d R$, and $R/\mathrm dR$ in Eqs.~(\ref{2_order0}), (\ref{2_order0_R+dR_weg}), and (\ref{2_order2}), respectively. }
\end{figure}

With (\ref{Riem_RNC}) we can already compute restrictions for the $\eta$~patch by employing (\ref{2)_general_summation_rule}) with $n=0$ and using terms $\mathcal{O}_g(\xi^2)$ to determine the condition. We will therefore denote the resulting maximal $\xi$ by $\xi^{(0)}_{\backslash R}$ with the adiabatic order of the truncated metric series once more given in the superscript and the terms we use to calculate the condition (i.e., the terms we drop), denoted by $\backslash R$ [instead of $\backslash (2)$ as in Sec.~\ref{sec: method}] in the subscript. Due to the diagonality of $\eta$ we see that $\text{max}_{\alpha,\beta}^\text{diag}\left\lbrace \left|\eta_{\alpha\beta}\right|\right\rbrace=1$, $\forall\,\alpha,\,\beta$, and the right hand-side of (\ref{2)_general_summation_rule}) is here consequently given by $\varepsilon$.

To determine the patch size, we then proceed iteratively as described in Sec.~\ref{subsec: RNC_applied}: First, we compute the patch size along the RNC axes. For that purpose, we set all $\xi$'s to $0$ except one $\xi^\mu$ and consider $\mathcal{O}_{g_{\alpha\beta}}(\xi^2)\!=\varepsilon$, $\forall \,\alpha,\beta$ with $\xi^\nu = 0$, $\forall\, \nu\!\neq\!\mu$ which gives the first set of conditions: $\xi^{\mu\,(0)}_{\backslash R} = \pm\sqrt{3}D\sqrt{\varepsilon}$, $\forall\,\mu$, where we defined $D= 2 M \left(r_0/2M\right)^{3/2} = r_0\sqrt{r_0/2M}$.

Second, we consider all possible $\xi^\mu$-$\xi^\nu$~combinations in $\mathcal{O}_{g_{\alpha\beta}}(\xi^2)$ with both other $\xi$'s set to $0$. Starting with the $\xi^0$-$\xi^1$~combination with $\xi^2=\xi^3=0$, we can see in Fig.~\ref{RNCpatch}a) that the domain of validity is here a square and the conditions along the $\xi^0$ and $\xi^1$~axes are thus valid for all combinations of the two coordinates. For the $\xi^0$-$\xi^2$~validity depicted in Fig.~\ref{RNCpatch}b), however, we see that the current conditions, which describe a square again, are insufficient, as for combinations of maximal $\xi^0$ and $\xi^2$ the error is larger than $\varepsilon$. The conditions must therefore be adjusted. Since, on the one hand, the correct description of this shape is very complicated, but, on the other hand, this patch is still squarelike, it is easiest to shrink the patch to a square with the diagonals. This yields new conditions for $\xi^0$ and $\xi^2$ given by $\xi^{0\,(0)}_{\backslash R} = \pm\sqrt{2}D\sqrt{\varepsilon}$ and $\xi^{2\,(0)}_{\backslash R} = \pm\sqrt{2}D\sqrt{\varepsilon}$. The combination of $\xi^0$ and $\xi^3$ produces an identical patch as in Fig.~\ref{RNCpatch}b), and we therefore also amend the $\xi^3$~condition to $\xi^{3\,(0)}_{\backslash R} = \pm\sqrt{2}D\sqrt{\varepsilon}$. The conditions derived from all other $\xi^\mu$-$\xi^\nu$~combinations in this second step are automatically satisfied given the adjusted $\xi$~restrictions.

Third, we examine the $\xi^\mu$-$\xi^\nu$~combinations again, but this time with only one other $\xi$ set to $0$ and the other bounded only by its maximal modulus value as determined above. A convenient patch of interest is shown in Fig.~\ref{RNCpatch}c). There, we see that the domain of validity for the $\xi^2$-$\xi^3$~combination with $\xi^0\neq 0$ and $\xi^1=0$ is a circle with radius $\sqrt{6D^2\varepsilon - 2(\xi^0)^2}$, and we add this condition to $\xi^{0\,(0)}_{\backslash R}$, $\xi^{2\,(0)}_{\backslash R}$, and $\xi^{3\,(0)}_{\backslash R}$. We find that thereby all other conditions are satisfied as well.

Fourth, the $\xi^\mu$-$\xi^\nu$~combinations are considered for the last time, now with neither of the other two $\xi$'s set to $0$, and we find no further adjustments to the conditions to be required.

In summary, Eq. (\ref{2)_general_summation_rule}) therefore gives the following conditions for the $\eta$~patch:
\begin{eqnarray}
\label{2_order0}
&& \left|\xi^{0\,(0)}_{\backslash R}\right| \leq \min \left\lbrace\sqrt{2} D \sqrt{\varepsilon},\sqrt{3 D^2 \varepsilon-\frac{(\xi^{2})^2 + (\xi^{3})^2}{2}}\right\rbrace\!, \nonumber\\
&& \left|\xi^{1\,(0)}_{\backslash R}\right| \leq \sqrt{3} D \sqrt{\varepsilon}\;, \nonumber \\
&& \left|\xi^{2\,(0)}_{\backslash R}\right| \leq \min \left\lbrace\sqrt{2} D \sqrt{\varepsilon},\sqrt{6 D^2 \varepsilon-(\xi^{3})^2 - 2(\xi^{0})^2 }\right\rbrace\!, \nonumber\\
&& \left|\xi^{3\,(0)}_{\backslash R}\right| \leq \min \left\lbrace\sqrt{2} D \sqrt{\varepsilon },\sqrt{6 D^2 \varepsilon-(\xi^{2})^2 - 2(\xi^{0})^2 }\right\rbrace\! .\quad\;\;
\end{eqnarray}
In the last two conditions we see the polar symmetry of these RNC which is a consequence of the spherical symmetry of a Schwarzschild black hole's geometry.

Note that the square roots in the conditions (\ref{2_order0}) can never become complex, because we can only plug in the maximal values of the other coordinates. Consider, for example, the RNC patch's boundary in the $\xi^2$-$\xi^3$~plane described $\xi^{2\,(0)}_{\backslash R} = \xi^{3\,(0)}_{\backslash R} = \pm\,\sqrt{2}D\sqrt{\varepsilon}$. The patch will then be restricted in the $\xi^0$~direction by $\sqrt{3 D^2 \varepsilon- 1/2 ((\xi^{2})^2 + (\xi^{3})^2)} = D\sqrt{\varepsilon}$. The restriction for the $\xi^1$~direction is unaffected.

In order to translate this patch size given in RNC to PG coordinates, we need to choose some geodesics along which we wish to compute the patch size in PG coordinates.

First, we consider the geodesic of the freely infalling probe and of the PG observer with 4-velocity (\ref{vel_PG_freefall}). With our choice of vierbein (\ref{vierbein_PG}) we find for this geodesic the transformed velocity $\lambda^0 = 1$, $\lambda^\mu = 0$, $\forall\,\mu\neq 0$. Using the conditions (\ref{2_order0}) we can now compute the maximal eigentime $\tau^{(0)}_{\text{PG}} = \max\left\lbrace \left|\xi^{0\,(0)}_{\backslash R}\right|\right\rbrace = \sqrt{2}D\sqrt{\varepsilon}$. The reparametrization of the curve with the observer's eigentime is here simply $t_{\text{obs}} = t_{\text{PG}}=\tau$. Therefore, we can integrate (\ref{vel_PG_freefall}) and plug $\tau^{(0)}_{\text{PG}}$ directly into $r(\tau)$. Thus, we find the minimal radial value of the PG observer's geodesic for which the $\eta$~patch as given by (\ref{2)_general_summation_rule}) is still valid
\begin{equation}
\label{rmin_PG_2_order0}
r_{\text{min}} = r\!\left(\tau^{(0)}_{\text{PG}}\right) = r_0\left(1-\frac{3}{\sqrt{2}}\sqrt{\varepsilon}\right)^{\frac{2}{3}}\;,
\end{equation}
with $r_0$ again the radial value of the reference point. We see that, for $r_0$ sufficiently close to $2M$ and sufficiently large $\varepsilon$, the RNC patch can cross the event horizon and extend into the black hole. This is due to the aforementioned regularity of RNC at the horizon. As an example, we take $r_0=2.1M$ with $\varepsilon = 0.01$ and find $r_{\text{min}}\approx 1.79M$, which is indeed inside the black hole. Also, in the limit $r_0\rightarrow \infty$ the patch size becomes arbitrarily large which reflects the asymptotic flatness of the Schwarzschild black hole's geometry.

As a second example we consider radially ingoing light rays. Since light travels on null geodesics, we parametrize the geodesic by the eigentime of the PG observer, i.e., the observer in question (as stated at the end of Sec.~\ref{sec: intro}). The 4-velocity then reads $v^{t_{\text{PG}}}=1$ and $v^r = -1-\sqrt{2M/r}$ and is transformed to $\lambda^0 = 1$ and $\lambda^1 = -1$ using (\ref{vierbein_PG}). The period of eigentime for which the RNC observer can describe ingoing light rays starting at the reference point $r=r_0$ is thus given by $\tau^{(0)}_{\text{li}} = \max\left\lbrace\left|\xi^{0\,(0)}_{\backslash R}\right|\right\rbrace = \sqrt{2}D\sqrt{\varepsilon}$.

Since, unfortunately, the relation $t^{\text{li}}_{\text{PG}}(r-r_0)$ obtained from integrating $v^r$ is not invertible in this case, we cannot directly quantify the minimal radius of validity for ingoing light rays. For a qualitative analysis of the patch size, we use the fact the time required for the PG observer to see ingoing light rays starting from the reference point reach the singularity at $r=0$, which is given by $T(r_0) = t^{\text{li}}_{\text{PG}}(-r_0) = r_0 - 2\sqrt{2Mr_0} + 4M\ln\!\left(1+\sqrt{r_0/(2M)}\right)$, remains finite for all finite $r_0$. As a consequence, there exist combinations of $r_0$ and the precision $\varepsilon$, for which ingoing light rays cross the horizon within the finite time $\tau^{(0)}_{\text{li}}$ provided by the RNC patch.

From these two examples we see that such RNC patches exist, that the subset of the causal future of $p$ covered by the patch after translation to another observer is large enough for it to describe physical objects crossing the event horizon. This is only possible for observers who see the causal future of $p$ reach into the black hole in their global $x$ coordinates, however. Since such RNC patches exist, normal coordinates can describe physics across the horizon and this description is accessible to other observers. Questions regarding the conservation of the causal structure at and across the horizon will be discussed in Sec.~\ref{subsec: causailty}.

It is important to note however, that when setting up RNC patches close to or even within the black hole, we do so in the presence of a quickly increasing background curvature. As we discussed in Sec.~\ref{sec: method} in our comments on Steps~1 and 2, this might cause problems if we take too few derivatives of the metric into account and/or choose the maximally allowed error $\varepsilon$ too large. In case of the $\eta$~patch as discussed above, we see this in two ways. First, from (\ref{rmin_PG_2_order0}) we read off that for $\varepsilon = 2/9$ the minimal radius of validity reaches $r_{\text{min}}=0$. Second, we solve $T(r_0=~2.1M)~\leq \tau^{(0)}_{\text{li}}$ which gives $\varepsilon \gtrsim 0.07$. We can therefore deduce that the $\eta$~patch as given by (\ref{2_order0}) has an upper bound on the error of $\varepsilon< 0.07$, as larger $\varepsilon$ would imply that we could, with only a small error, describe light  rays or even the PG observer falling into the singularity using a flat Minkowski metric, which is unreasonable.

To set up such RNC patches close to the black hole's singularity, we are thus required to take higher orders of the metric expansion into consideration, either by calculating some $g^{(n)}$~patch instead of the $\eta$~patch or by calculating the $\eta$~patch size using (\ref{2)_general_extended_summation_rule}) rather than (\ref{2)_general_summation_rule}). In the above case of the Minkowski patch, we will calculate the $\eta$~patch size using (\ref{2)_general_extended_summation_rule}) with $k=3$. For that purpose, we first need to compute the derivatives of the Riemann tensor in RNC which occur in $\mathcal{O}_g(\xi^3)$ and which we therefore take into account in determining the $\eta$~patch size. Some computational effort is required to obtain these derivatives which are, by means of the coordinate transformation (\ref{RNC_coord_trafo}), given by
\begin{eqnarray}
\label{Riem_RNC_deriv}
&& R_{\alpha\beta\gamma\delta,\mu} = e^a_\alpha e^b_\beta e^c_\gamma e^d_\delta e^m_\mu \left(R_{abcd,m} - \Gamma^n{}_{ma}R_{nbcd}  \right.\nonumber \\
&& \left.  - \Gamma^n{}_{mb}R_{ancd} - \Gamma^n{}_{mc}R_{abnd} - \Gamma^n{}_{md}R_{abcn}\right).
\end{eqnarray}
The explicit terms resulting from this are given in Appendix~\ref{sec: appendix b}.

We plug (\ref{Riem_RNC}) and (\ref{Riem_RNC_deriv}) into (\ref{2)_general_extended_summation_rule}) with $k=3$ and obtain instead of (\ref{2_order0}) improved $\eta$~patch conditions: \pagebreak
\begin{eqnarray}
\label{2_order0_R+dR_weg}
&& \left|\xi^{0\,(0)}_{\backslash \mathrm d R}\right| \leq \sqrt{2} D \sqrt{\varepsilon}-\frac{5}{2} D \varepsilon\;, \nonumber\\
&& \left|\xi^{1\,(0)}_{\backslash \mathrm d R}\right| \leq \sqrt{3} D \sqrt{\varepsilon}-\frac{9}{4} \sqrt{\frac{r_0}{2M}} D \varepsilon\;, \nonumber\\
&& \left|\xi^{2\,(0)}_{\backslash \mathrm d R}\right| \leq \min \Bigg\lbrace\sqrt{2} D \sqrt{\varepsilon}-\frac{5}{2} D \varepsilon,\nonumber\\
&&\frac{\sqrt{12 D^3 \varepsilon-(\xi^{3})^2 \left(9 \xi^{0}+2 D\right)-2 (\xi^{0})^2 \left(3 \xi^{0}+2 D\right) }}{\sqrt{9 \xi^{0}+2 D}}\Bigg\rbrace  .\quad\;\;
\end{eqnarray}
We label these conditions by $\backslash\text{d}R$, because now the coefficients of the highest order terms taken into account for calculating the patch size are given by first derivatives of the Riemann tensor. Note that since the full condition terms are very lengthy here, we Taylor expanded all of them in $\varepsilon$, except for the directional dependence term. Notice, these conditions are again real, if the smallness of $\varepsilon$ is respected. Furthermore, the $\xi^3$~condition can be obtained by symmetry from the $\xi^2$~restriction by interchanging $\xi^2$ with $\xi^3$. Finally, we also have a condition for the directional dependence of $\xi^0$ which follows from solving the $\xi^2$ or $\xi^3$~condition for $\xi^0$. For reasons of clarity and comprehensibility, we omitted this $\xi^0$~condition in the above.

\begin{figure}
	\includegraphics[scale=1]{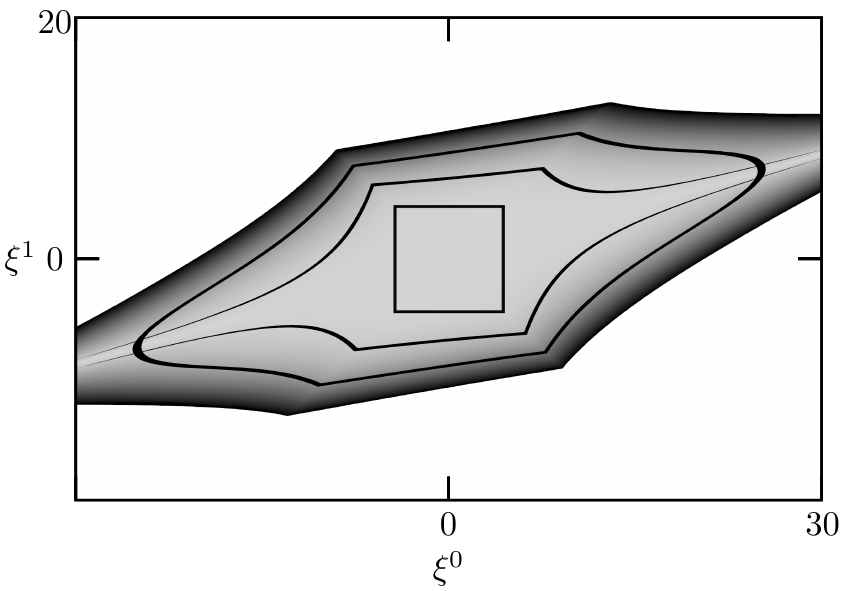}%
	\caption{\label{RNCpatch3rd4th} Depicted are the boundaries of the square shaped Minkowksi patch in the middle and of the second and third order RNC patches encircling it with $M=1$ and $r_0=24$. These boundaries correspond to an error of $\varepsilon\in[\,0.9\times 10^{-3},10^{-3\,}]$. The fourth order patch with maximal error $\varepsilon=10^{-3}$ is shown completely. The change in size and shape of the patches with increasing order is complicated, but a qualitative increase can be seen.}
\end{figure}

To produce reasonable functional dependences in these conditions, we again adjusted the patches found in the iteration procedure in the simplest, yet most sensible, way. As an example, consider the domain of validity shown in Fig.~\ref{RNCpatch}d), especially the diagonal ``arms'' of the patch which reach all the way to infinity. This would correspond to a noncompact, open domain of validity, which is unreasonable. Thus, we cut off these ``arms'' and again fit a square into the central region of the patch. Truncating the ``arms'' is unproblematic, however, as we will explain in detail in the discussion of Fig.~\ref{RNCpatch3rd4th}.

Above we have seen that the conditions (\ref{2_order0}) gave physically unreasonable results like radially ingoing light rays reaching the singularity for errors $\varepsilon\gtrsim 0.07$, which therefore marked the upper bound for the error. The conditions (\ref{2_order0_R+dR_weg}) contain more curvature information and will therefore both allow for a larger maximal error and improve the patch size. Employing (\ref{2_order0_R+dR_weg}) in analogous calculations for the freely infalling observer as for (\ref{rmin_PG_2_order0}), we now obtain a minimal radial value of validity
\begin{equation}
\label{rmin_PG_2_order0_R+dR_weg}
r_{\text{min}} = r_0\left(1-\frac{3}{\sqrt{2}}\sqrt{\varepsilon}+\frac{15}{4}\varepsilon\right)^{\frac 2 3}.
\end{equation}
For $r_0=2.1M$ and $\varepsilon = 0.01$, this gives an increased $r_{\text{min}}\approx 1.85M$. The reason for $r_{\text{min}}$ determined by (\ref{rmin_PG_2_order0_R+dR_weg}) being larger as the result of (\ref{rmin_PG_2_order0}) is that taking higher orders of the metric series into account improves the domain of validity regarding the accuracy of describing the background, but does not necessarily increase it. For the Schwarzschild metric, the curvature grows quickly close to the singularity, so taking Riemann tensor derivatives into account will actually decrease the patch size for small $r_0$ compared to when they are ignored. We also see this improvement by noting that $r_{\text{min}}=0$ is impossible in (\ref{rmin_PG_2_order0_R+dR_weg}). Furthermore, $r_{\text{min}}$ given by (\ref{rmin_PG_2_order0_R+dR_weg}) decreases only until $\varepsilon = 0.08$, after which it grows. Analogously, $\tau^{(0)}_{\text{li}}$ increases only until $\varepsilon = 0.08$ and decreases for larger $\varepsilon$. We have therefore increased the upper bound on the maximal error to $\varepsilon < 0.08$, as only after that we see unreasonable behavior. We will provide a detailed comparison of the different patch sizes produced by our method at the end of our discussion on the patch size for the PG observer.

We can now also quantitatively compare the $\eta$~patch size (\ref{2_order0_R+dR_weg}) resulting from our procedure with the estimate from the literature (\ref{patch_size_lit}). Plugging (\ref{Riem_RNC}) and (\ref{Riem_RNC_deriv}) into (\ref{patch_size_lit}) we obtain $\tau\ll\text{min}\{r_0/3\, , r_0/3 \sqrt{r_0/(2M)}\}$, which yields a spherical patch in RNC with radius $\ll r_0/3$ outside the black hole and radius $\ll r_0/3\sqrt{r_0/(2M)}$ inside. Comparing this with (\ref{2_order0_R+dR_weg}), we see that this estimate is too restricting, since in (\ref{2_order0_R+dR_weg}) we have a leading order term $\propto\sqrt{\varepsilon}$, and it also lacks the complicated directional dependence. If we consider the above example of $r_0=2.1M$, $\varepsilon=0.01$ for the literature estimate by plugging $\tau=\varepsilon r_0/3 = 7\times 10^{-3}M$ into the PG observer's geodesic, we find a minimal radial value of approximately $2.08M$. In contrast to our result $r_{\text{min}}\approx 1.85M$, the point of the minimal radial value for the freely falling observer estimated by the literature is still outside the black hole. Also, plugging $r=2M$ into $t^{\text{li}}_{\text{PG}}(r-r_0)$, which describes the PG observer's eigentime required for radially ingoing light rays to reach the radius $r$, we find $t^{\text{li}}_{\text{PG}}(2M-2.1M) \approx 0.05M > \tau$. The literature therefore estimates the RNC patch so small that the subset of the causal future of $p$ covered by the RNC patch does not range across the horizon. However, we have shown that it indeed does.

Instead of calculating the $\eta$~patch using (\ref{2)_general_extended_summation_rule}), we can also take higher orders of the metric expansion into account by calculating some $g^{(n)}$~patch. Therefore, let us also present the conditions for the patch covered by $g^{(2)}(\xi)$, which are found by employing our results from Sec.~\ref{subsec: RNC_applied}. Plugging (\ref{Riem_RNC}) and the results of (\ref{Riem_RNC_deriv}) into (\ref{2)_general_order2}), which was obtained by employing (\ref{2)_general_summation_rule}), we obtain the $g^{(2)}$~conditions
\begin{eqnarray}
\label{2_order2}
\left|\xi^{0,1\,(2)}_{\backslash \mathrm d R}\right| &\leq & \left(\frac{2}{{\sqrt{d}+1}}\right)^\frac{1}{3} D{\varepsilon }^{\frac{1}{3}}+\frac{8}{9}\frac{ D }{ {\sqrt{d}}+1} \varepsilon\;, \nonumber\\
\left|\xi^{2\,(2)}_{\backslash \mathrm d R}\right| &\leq & \min \Bigg\lbrace \left(\frac{4  }{{3}\sqrt{d}}\right)^\frac{1}{3} D {\varepsilon}^\frac{1}{3}-\frac{8  }{27}\frac{ D }{\sqrt{d}} \varepsilon,\nonumber\\
&&\, M d  \frac{6 D^2 + (\xi^{1})^2 - 2(\xi^{3})^2}{3 \xi^{1} \xi^{3}} \varepsilon \Bigg\rbrace\;,
\end{eqnarray}
where we additionally defined $d = r_0/2M$. Note that as before we Taylor expanded the conditions in $\varepsilon$. The $\xi^0$ and $\xi^1$~conditions are here equal except for an additional directional $\xi^1$~dependence which we again find by solving the $\xi^2$~condition for $\xi^1$. Also, the $\xi^3$~condition is once more obtained from the $\xi^{2}$~condition as described above. Since the general conditions for $\xi^2$ and $\xi^3$ are extremely lengthy and complicated, we restrict ourselves to the case of $\xi^0$ and $\xi^1$ with an equal sign which yields the short conditions~(\ref{2_order2}). For the conditions of the full patch we refer to our \textit{Mathematica} code \cite{hoegl_bruno_2020_3966891}.

To observe the growth of the patch size achieved by including curvature corrections into the metric, we again plug $\xi^{0\,(2)}$ into the PG observer's geodesic for $r_0=2.1M$, $\varepsilon=0.01$ and find $r_{\text{min}} \approx 1.61M$. The patch now reaches further into the black hole.

After having discussed different patch sizes determined by different sets of conditions, we want to put them in relation. For that, compare first Fig.~\ref{RNCpatch}a) with Fig.~\ref{RNCpatch}d) and note the discrepancies in both shape and size of the $\eta$~patches determined first by dropping only $\mathcal{O}_g(\xi^2)$ and second by dropping $\mathcal{O}_g(\xi^2)+\mathcal{O}_g(\xi^3)$. Note especially that the patch in Fig.~\ref{RNCpatch}d), where higher orders of the metric series are taken into account, is actually smaller than the one in Fig.~\ref{RNCpatch}a) except for the pathological arms. This is because, as we discussed earlier, taking higher orders of the metric series into account improves the patch size but does not need to increase it. We can also see this by further comparing these patch sizes with the $g^{(2)}$~patch's $\xi^0$-$\xi^1$~validity given in (\ref{2_order2}) and depicted in Fig.~\ref{RNCpatch}e) which is even smaller. The reason for this is that, given a reference point at $r_0=24M$ as in Fig.~\ref{RNCpatch}, an error of $\varepsilon=0.1$ is too large.

If we reduce the error to $\varepsilon=10^{-3}$, we find instead of Fig.~\ref{RNCpatch}a) for the $\eta$~patch with $k=2$ the patch in Fig.~\ref{RNCpatch}f), instead of Fig.~\ref{RNCpatch}d) for the $\eta$~patch with $k=3$ the domain in Fig.~\ref{RNCpatch}g) and instead of the $g^{(2)}$~patch in Fig.~\ref{RNCpatch}e) the one in Fig.~\ref{RNCpatch}h). Comparing these patches, we see that Fig.~\ref{RNCpatch}f) and Fig.~\ref{RNCpatch}g) now almost agree, with the patch in Fig.~\ref{RNCpatch}g) being slightly larger, as expected. We can see this also by noting that the conditions of (\ref{2_order0}) and (\ref{2_order0_R+dR_weg}) in the $\xi^0$-$\xi^1$~plane agree in the limit $\varepsilon\rightarrow 0$, as the higher order terms in (\ref{2_order0_R+dR_weg}) become strictly irrelevant. Furthermore, the $g^{(2)}$~patch in Fig.~\ref{RNCpatch}h) is now substantially larger than both $\eta$~patches in Figs.~\ref{RNCpatch}f) and \ref{RNCpatch}g). Since this reflects the expected behavior of a Taylor series, where including higher orders of the expansion increases the domain of validity, we deduce that $\varepsilon = 10^{-3}$ is a better choice for the error than $\varepsilon = 0.1$.

Additionally, we want to discuss the case of considering the metric together with the Riemann tensor. For that purpose, we use the metric $g^{(2)}$ and again our results from Sec.~\ref{subsec: RNC_applied}. Specifically, we insert (\ref{Riem_RNC}) and (\ref{Riem_RNC_deriv}) into (\ref{4)_general_Riem_1}) and (\ref{4)_general_Riem_2}) and thus find the patch size regulations
\begin{eqnarray}
\label{4_order2}
&\left|\xi^{0\,(2)}_{\mathrm{Riem}}\right| \leq & \min \Bigg\lbrace \frac{3 D }{\sqrt{2}} \sqrt{\varepsilon},  \frac{D}{3} \varepsilon \Bigg\rbrace\;, \nonumber\\
&\left|\xi^{1\,(2)}_{\mathrm{Riem}}\right| \leq & \min \Bigg\lbrace \frac{3D}{\sqrt{2}} \sqrt{\varepsilon} , \frac{r_0}{3}\varepsilon  \Bigg\rbrace\;,\nonumber\\
&\left|\xi^{2\,(2)}_{\mathrm{Riem}}\right| \leq & \min \Bigg\lbrace \frac{3D}{\sqrt{2}} \sqrt{\varepsilon} , \frac{r_0}{3}\varepsilon ,  
\sqrt{9 r_0^2 d \varepsilon -2 (\xi ^1)^2-2 (\xi^{3})^2},\nonumber \\
&& \sqrt{18 r_0^2 d \varepsilon - 4(\xi^0)^2 - 2(\xi^1)^2 -(\xi^{3})^2}\Bigg\rbrace\;.
\end{eqnarray}

\pagebreak
Calculating the minimal radial value of validity for the PG observer using (\ref{4_order2}) with $r_0=2.1M$ and $\varepsilon=0.01$, we find $r_{\text{min}}\approx 2.09M$ and see that, as discussed in our comments on Step 4 in Secs.~\ref{sec: method} and \ref{subsec: RNC_applied}, the domain of validity has indeed decreased in size drastically.

Finally, let us observe the further growth of the validity domain for metric expansions up to higher orders $n=3$ and $n=4$ compared to the $\eta$ and $g^{(2)}$~patch described by (\ref{2_order0}) and (\ref{2_order2}), respectively. Since the conditions for the $g^{(3)}$ and $g^{(4)}$~patch are extremely lengthy, however, we abstain from writing them down and instead show the growth of the patch size graphically. For that purpose, we plot in Fig.~\ref{RNCpatch3rd4th} the $g^{(4)}$~patch in the $\xi^0$-$\xi^1$~plane for $\varepsilon=10^{-3}$ obtained using (\ref{2)_general_summation_rule}). Figure~\ref{RNCpatch3rd4th} additionally shows the boundary regions of the $\eta$~patch from Fig.~\ref{RNCpatch}f) and the $g^{(2)}$~patch from Fig.~\ref{RNCpatch}h) as well as of the $g^{(3)}$~patch described by the error interval $\varepsilon\in[\,0.9\times10^{-3},10^{-3\,}]$. We also employed (\ref{2)_general_summation_rule}) for the $g^{(3)}$~patch.

The smallest and the second smallest patch are the two familiar patches from Figs.~\ref{RNCpatch}f) and \ref{RNCpatch}h). The patch next in size corresponds to the metric expansion $g^{(3)}$ and the largest patch describes the domain of validity for $g^{(4)}$. All in all, we see a continuous growth of the patch sizes with increasing adiabatic order $n$. The thickness of the boundary lines for the patches with $n=0, 2, 3$ depicts how fast the error grows for the respective patches: the thicker the boundary, the slower the error increases.

It is important to note that while the arms reaching to infinity in the $g^{(2)}$~patch disappear for the $g^{(3)}$~patch, they reoccur for $g^{(4)}$. These arms are formed along lines describing $\xi^0$-$\xi^1$~configurations for which both the first and the third derivatives of the Riemann tensor vanish such that the coefficients of $\mathcal{O}_g(\xi^3)$ and $\mathcal{O}_g(\xi^5)$ vanish as well. If we calculate the $g^{(2)}$ and $g^{(4)}$~patch sizes using (\ref{2)_general_summation_rule}), we could therefore assume the validity to reach to infinity. It is, however, safe to ignore such pathological arms, because taking into consideration higher orders when calculating the patch sizes, namely by employing (\ref{2)_general_extended_summation_rule}) with $k\geq 2$ instead of (\ref{2)_general_summation_rule}), erases these arms. This can be seen by considering the $g^{(3)}$~patch. It is calculated using only (\ref{2)_general_summation_rule}), but since the coefficient of $\mathcal{O}_g(\xi^4)$ also depends on nonderivative terms of the Riemann tensor, this coefficient is finite along the lines and thus the arms are cut off. Analogously, higher order terms of even $n$ do not depend solely on derivatives of the Riemann tensor and taking them into account when calculating, for example, the  $g^{(4)}$~patch, will cut the arms.

\subsection{Schwarzschild observer}
\label{subsec: SS}

In the case of the PG observer the (full) RNC were simply another coordinate system associated with this observer. Let us now derive the patch size for the Schwarzschild observer as an example of an observer for whom the inside of the black hole is excluded from the causal future of the outside. In order to construct a vierbein corresponding to the coordinate transformation $\{ t, r, \theta, \phi \}$ $\rightarrow$ $\{\xi^0,\xi^1,\xi^2,\xi^3\}$, one can proceed analogously to (\ref{vierbein_PG}), setting $e^a_0 = v^a$ with $v_a$ given in (\ref{vel_SS_freefall}) and fixing the other components by orthonormality $g_{ab} e^a_\alpha e^b_\beta = \eta_{\alpha\beta}$, which yields
\begin{eqnarray}
\label{vierbein_SS}
&& e^a_0 = v^a, \quad e^t_1 = v^t v^r, \quad e^r_1 = 1, \nonumber \\
&& e^\theta_2 = \frac{1}{r} , \quad e^\phi_3 = \frac{1}{r \sin\theta}\;.
\end{eqnarray}

With this vierbein one tensor transforms the Riemann tensor and finds the same components as given in (\ref{Riem_RNC}). Also, calculating Riemann tensor derivatives according to (\ref{Riem_RNC_deriv}) gives again the same terms as in the PG case (see Appendix~\ref{sec: appendix b}). As we discussed above, this is because Schwarzschild and PG coordinates both describe the same geometry of a Schwarzschild black hole. Note also that the curvature remains finite at the horizon $r=2M$ and the RNC obtained from Schwarzschild coordinates are also regular, which, as we discussed above, reflects the choice of the RNC observer to parametrize each geodesic using its respective eigentime.

Consequently, employing Step~2 with (\ref{2)_general_summation_rule}) for $n=0$ and $n=2$ as well as with (\ref{2)_general_extended_summation_rule}) for $n=0$ and $k=3$ and also Step~4 with $n=2$ for the Riemann tensor yields the same RNC conditions as given in (\ref{2_order0}), (\ref{2_order2}), (\ref{2_order0_R+dR_weg}), and (\ref{4_order2}). 

The translation of these patch sizes to the Schwarzschild observer is more involved than for the PG observer, however. The reason for this is that the PG observer uses a clock which is more adapted to RNC than the Schwarzschild observer. To see this in detail, we will again consider the two examples of the freely infalling probe and of ingoing light rays.

We begin again by considering the freely falling probe with 4-velocity (\ref{vel_SS_freefall}). With our choice of vierbein (\ref{vierbein_SS}) we again have for this geodesic $\lambda^0=1$ and $\lambda^\mu=0$, $\forall\,\mu\neq 0$ and thus find the same maximal proper lengths $\tau^{(n)}_{\text{pr}}$ as given in Sec.~\ref{subsec: PG}. Furthermore, integrating $v^r$ yields the same $r(\tau)$ and therefore the same minimal radii $r_{\text{min}}$ as in our calculations for the PG observer.

Now we have to reparametrize the curve by the Schwarzschild observer's eigentime $t_{\text{obs}} = t(\tau)$, however. As was the case for light rays and the PG observer, $t(\tau)$ is not invertible here and we cannot directly calculate the minimal radius of RNC validity for the probe as seen by the Schwarzschild observer by plugging $t^{(0)} = t(\tau^{(n)}_{\text{pr}})$ into $r(\tau)$. Therefore, we again analyze the patch qualitatively.

For this purpose, we can use the fact that the time needed to reach some $r\geq 2M$ diverges as the event horizon is approached: $t(r)\propto -2M\ln\!\left(1-\sqrt{2M/r}\right)$ for $r\rightarrow 2M$. In Sec.~\ref{subsec: PG} we have seen, however, that the RNC observer can see their patch crossing the horizon, namely that for reference points sufficiently close to $2M$ and large enough $\varepsilon$ we can have $r_{\text{min}} = r(\tau^{(n)}_{\text{pr}})\leq 2M$. For $r_{\min}>2M$, we determine the temporal validity of the RNC patch along this geodesic as seen by the Schwarzschild observer by plugging $r_{\min}$ into $t(r)$. Consequently, this temporal validity diverges as soon as the RNC observer sees their geodesic reaching the horizon, i.e., $r_{\text{min}}= 2M$, and the horizon is reached within the RNC patch in the limit of infinite time $t$. Any further progress of the freely falling probe inside the black hole remains hidden from the Schwarzschild observer, however, as $r_{\min}<2M$ cannot be plugged into $t(r)$ and the temporal validity has already grown to infinity. We also need to note that such temporally infinite validities only hold ``in the direction'' in which geodesics cross the horizon. For example, the temporal validity in the past of the freely falling probe is finite.

It is important to verify that the subset of the causal future of the RNC reference point covered by the RNC patch does not cross the black hole horizon for the Schwarzschild observer. Therefore, we again consider radially ingoing light rays. As before, we parametrize light by the eigentime of the observer in question. The patch size along the light ray geodesic is determined by how long the RNC observer can describe the light ray using their eigentime. The observer in question for the parametrization of the light ray geodesic is therefore the PG observer, who uses the same clock as the RNC observer, and not the Schwarzschild observer. Thus, we have to determine the length of Schwarzschild time $t$ that corresponds to the time $t_{\text{PG}}$ for which the light rays remain in the RNC patch as seen by the PG observer. For that, we first deduce from (\ref{line_elem_SS}) the ingoing light ray's coordinate velocity $\mathrm d r/\mathrm d t = -f(r)$ in Schwarzschild coordinates. Second, we have to reparametrize the light's geodesic by $t_{\text{PG}}$. Using (\ref{proper_time}) with $\mathrm d r/\mathrm d t = -f(r)$ we\linebreak obtain $\text d t/\text d t_{\text{PG}} = f^{-1}(r)(1+\sqrt{2M/r})$ as well as $\mathrm d r/\mathrm d t_{\text{PG}} = -1-\sqrt{2M/r}$. This velocity can now be tensor transformed with (\ref{vierbein_SS}) which finally yields $\lambda^0=1$ and $\lambda^1=-1$ as expected for ingoing light rays. The length of time for which the RNC observer can describe ingoing light rays is thus given by $\tau^{(n)}_{\text{li}}$ from Sec.~\ref{subsec: PG}.

Integration of $\mathrm d r/\mathrm d t$ shows again the infinite time required for the Schwarzschild observer to see the light rays reach the horizon $t(r)\propto -2M\ln\!\left(f(r)\right)$ for $r\rightarrow 2M$. From Sec.~\ref{subsec: PG} we already know, however, that the RNC observer can see light rays crossing the event horizon; i.e., $r(\tau^{(n)}_{\text{li}})\leq 2M$ is possible in PG coordinates. By the same reasoning as for the freely falling probe, we see in such cases again temporally infinitely valid RNC patches which allow the Schwarzschild observer to see the light rays reach the horizon within the patches in the limit of infinite $t$.

The Schwarzschild coordinates are an example of coordinate systems that break down at the horizon. As a consequence, the causal subset of the RNC patch translated to Schwarzschild coordinates must not cross the horizon. Transforming the patch size from the RNC observer to the Schwarzschild observer and taking the latter's use of their eigentime as a profoundly different clock into account, we find this essential demand fulfilled. The structure of the RNC reference point's causal future in Schwarzschild coordinates is preserved.

Schwarzschild coordinates are also an example of coordinate systems which show the importance of our discussion in the comment for Step~5 concerning the relevance of using the correct parametrization when translating the patch size to other observers. If we had simply plugged the RNC patch size conditions into the coordinate transformation (\ref{RNC_coord_trafo}), we would erroneously have deduced that for the Schwarzschild coordinates the causal future of the reference point covered by the RNC patch crosses the horizon. We would have found the same result, if we had considered $r(\tau)$ for the freely falling probe or the ingoing light rays and not reparametrized these geodesics by $t$. This is because in both cases we would have obtained the patch size as seen by the RNC observer and not the Schwarzschild observer. Only after this reparametrization did we find the real translated patch size as seen by the Schwarzschild observer.

This analysis of the PG and Schwarzschild observer as examples of observers that do and do not see horizon crossing of RNC patches, respectively, can also be used to determine whether FNC or FNCP can cross the event horizon. Only FNC and FNCP set up around the geodesics of observers who see horizon crossing themselves, e.g., infalling observers, will reach inside the black hole. FNC and FNCP constructed around other observers, for example, orbiting ones, cannot cross the horizon, just as was the case for the Schwarzschild observer.

\subsection{Causality at the event horizon}
\label{subsec: causailty}

Having computed that RNC patches can cross the black hole horizon, it is crucial that we investigate the causal structure in such patches. This means, we require that in RNC physical systems can only cross the horizon from the outside to the inside and that it is impossible for an observer to interact with systems inside the black hole as long as this observer is outside the black hole.

The event horizon for an eternal black hole equals its apparent horizon \cite{Faraoni:2015ula}. The latter allows, in contrast to an event horizon, for a description in a finite spacetime region and is thus the preferred object for analyzing the local causal structure. For spherically symmetric spacetimes, apparent horizons are characterized as the null hypersurfaces across which radially outward directed light rays change the sign of their coordinate velocity. For a Schwarzschild spacetime in PG coordinates this velocity is computed to be $(1,1-\sqrt{2M/r},0,0)$. Therefore, the change of sign occurs at $r=2M$ with velocity $(1,0,0,0)$. This uniquely determines the location of the apparent horizon. Having identified the radial location of the horizon, we can alternatively specify it as the point that is reached by the PG observer after the eigentime $\tau_{\text{PG}}(r_0) = 2/3(D-2M)$ measured from the reference point at $r=r_0$. Since the apparent horizon equals the event horizon in our system, we use from now on again the latter term for convenience.

To investigate causality at the horizon, it is sufficient that we consider an RNC patch with solely the leading Minkowski part around a reference point outside of, yet sufficiently close to, the horizon, such that an infinitesimal neighborhood of the reference point already covers part of the horizon. In Sec.~\ref{subsec: PG} we showed that $\eta$~patches can indeed cross the horizon.

In order to translate the horizon in PG coordinates as described above to RNC, we proceed as follows: First, we transform the velocity $(1,0,0,0)$ with the vierbein (\ref{vierbein_PG}) at the reference point, which yields $\lambda^0=1$ and $\lambda^1=\sqrt{2M/r_0}$. Second, we shift the resulting line such that it intersects the $\xi^0$~axis at $\xi^0=\tau_{\text{PG}}$. All in all, in RNC the radial evolution of the horizon along the PG observer's geodesic is described by the straight line $\Omega(\xi^0) = (\xi^0,\sqrt{2M/r_0}(\xi^0-\tau_{\text{PG}}),0,0)$. For this construction to hold, it is in fact required that the reference point is infinitesimally close to the horizon, as we will show shortly. We therefore have $r_0\rightarrow 2M$ and $(\mathrm{d}_{\xi^0} \Omega)^\alpha\rightarrow(1,1,0,0)$.

\begin{figure}
	\includegraphics[scale=1]{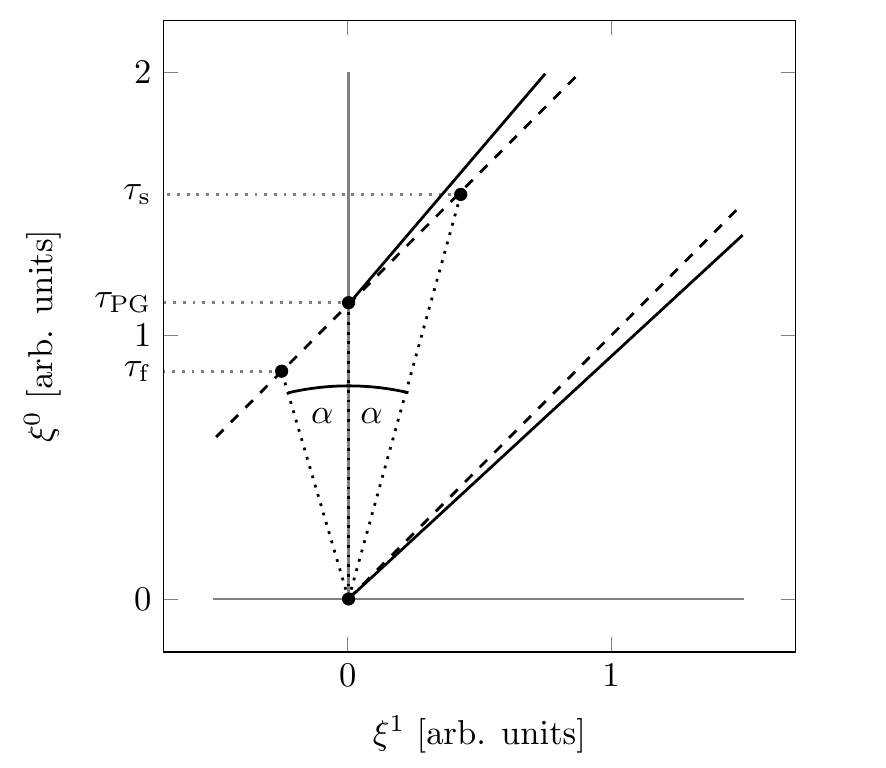}
	\caption{\label{figRNCgeod} Geodesics for radially outward directed light rays in an RNC patch of a Schwarzschild geometry with some mass $M$. The dashed lines represent the outgoing null geodesics in Minkowski spacetime $\xi^0 = \xi^1 + \text{const}$. The dotted lines correspond to three timelike observers following radially infalling geodesics with different initial velocities solved in PG coordinates. The solid lines are outward directed null geodesics solved in RNC up to second adiabatic order.}	
\end{figure}

If the RNC expansion point is set outside the black hole $r_0\gtrsim 2M$, the horizon is given by the upper dashed line in the spacetime diagram in Fig.~\ref{figRNCgeod}, which here is at $45^{\circ}$, but is to be understood as infinitesimally steeper, since we have $(\mathrm{d}_{\xi^0} \Omega)^1 = \sqrt{2M/r_0} \lesssim 1$. Outgoing light rays emitted at the reference point therefore follow the lower dashed line at $45^{\circ}$ and diverge radially from the horizon to the outside. For negative $\xi^0$ we encounter a crossing point of the horizon with the outgoing light rays which is pathological, since it would correspond to light rays crossing the horizon from the inside to the outside. This indicates the breakdown of the parallel shift construction at this crossing point; for large $\xi^0$, the horizon is no longer represented by the parallely shifted line which describes it well for small $\xi^0$. Note, however, that since $(\mathrm{d}_{\xi^0} \Omega)^1 \rightarrow 1$ for $r_0 \rightarrow 2M$, the intersection then occurs at $\xi^0\rightarrow -\infty$. For a reference point close to the horizon, the breakdown of the parallel shift therefore occurs far outside any domain of validity.

For an RNC expansion at the event horizon $r_0=2M$, the horizon is given by the lower dashed line at $45^{\circ}$ in Fig.~\ref{figRNCgeod}, and we find outgoing (or rather outward directed) light rays emitted at the origin to remain on the horizon. This reflects the aforementioned characterization of the event horizon as a null hypersurface for outgoing light rays. Furthermore, we see that any other lightlike geodesic pointing radially outward starting with $\xi^1<0$ will not cross the event horizon and thus remains inside the black hole.

Finally, if we chose the reference point inside the black hole $r_0\lesssim 2M$, the horizon crosses the $\xi^0$ axes at negative $\xi^0$ and infinitesimally less steep than $45^{\circ}$ [now we have $(\mathrm{d}_{\xi^0} \Omega)^1 \gtrsim 1$]. Therefore, all light rays emitted from the reference point will fall into the black hole, even outward directed light rays following the line $(\xi^0,\xi^0,0,0)$, which seem to be outgoing to the RNC observer. For negative $\xi^0$ we once more see such light rays crossing the horizon, which would correspond to outward directed light rays in the black hole having crossed the horizon from the outside at some time in the past. This is again pathological, as these light rays start infinitesimally close to the horizon from inside for the infinite past. They do not cross the horizon from the outside. As before, the parallely shifted line only describes the horizon for sufficient small $\xi^0$.

Let us also check whether the causal order of future and past events remains intact in the RNC with the following example: We consider two additional timelike observers on radial geodesics starting at the reference point outside the black hole. The first observer starts with an inwards radial velocity $v^r=V_{\text{f}} \in\; ]-\infty,-\sqrt{2M/r_0}\,[$ faster than that of the freely falling PG observer and the second one with a slower one $v^r=V_{\text{s}} \in\; ]-\sqrt{2M/r_0},0\,]$. Using the vierbein (\ref{vierbein_PG}) we transform $v^r$ together with the corresponding $v^{t_{\text{PG}}}$ computed with (\ref{line_elem_PG}) and obtain the velocity in RNC
\begin{eqnarray}
\label{vel_causality_RNC}
&& \lambda^0(V) = \frac{\sqrt{\frac{2M}{r_0}}V + \sqrt{V^2 + f(r_0)}}{f(r_0)}\;, \nonumber\\
&& \lambda^1(V) = \frac{V + \sqrt{\frac{2M}{r_0}}\sqrt{V^2 + f(r_0)}}{f(r_0)}\;,
\end{eqnarray}
for which we note that $\lambda^0(V)>1$, $\forall\, V,r$ as well as $\lambda^1(V_{\text{f}})<0$ and $\lambda^1(V_{\text{s}})>0$, $\forall\, r$. We can now check if the angles $\alpha(V_{\text{s}})$ and $\alpha(V_{\text{f}})$ formed by the lines $(\xi^0,0,0,0)$ of the PG observer and $(\xi^0\lambda^0(V),\xi^0\lambda^1(V),0,0)$ of the other observers are always smaller than $\pi/4$ which corresponds to a connection between these observers by causal curves. We find these two angles to be given by $\alpha(V_{\text{s}}) = \pi/2 - \tan^{-1}\!\left(\lambda^0(V_{\text{s}})/\lambda^1(V_{\text{s}})\right)$ as well as $\alpha(V_{\text{f}}) = \pi/2 + \tan^{-1}\!\left(\lambda^0(V_{\text{f}})/\lambda^1(V_{\text{f}})\right)$ and we therefore require $\lambda^0(V_{\text{s}})/\lambda^1(V_{\text{s}}) > 1$ and $\lambda^0(V_{\text{f}})/\lambda^1(V_{\text{f}}) < -1$, which we find satisfied for all possible $V_{\text{s}}$ and $V_{\text{f}}$, respectively. It now remains to verify that the faster observer reaches the horizon earlier than the PG observer and that the slower observer takes longer. The PG observer crosses the horizon after the eigentime $\tau_{\text{PG}}$. For simplicity, we assume the horizon to be at $45^{\circ}$ and demand $\tau_{\text{f}}:=\tau_{\text{PG}}+\lambda^1(V_{\text{f}})\tau_{\text{PG}} < \lambda^0(V_{\text{f}})\tau_{\text{PG}}$ for the faster observer and $\tau_{\text{s}}:=\tau_{\text{PG}} + \lambda^1(V_{\text{s}})\tau_{\text{PG}} > \lambda^0(V_{\text{s}})\tau_{\text{PG}}$ for the slower one. Again, we find these inequalities fulfilled by all possible $V_{\text{f}}$ and $V_{\text{s}}$ as given above and since the horizon is actually infinitesimally steeper, the effect is only increased.

Performing the same analysis for the three observers inside the black hole, with the event horizon substituted by the geodesic of outward directed light rays and also $V_{\text{s}} \in\; ]-\sqrt{2M/r_0},-\sqrt{-f(r_0)}\,]$, we find all of the demands to be satisfied again.

Above, we computed the geodesics going through the expansion point in the global PG coordinates and transformed them to RNC. There is no causality violation in PG coordinates and we found this to be translated to RNC. Now we want to investigate whether causality violation occurs in RNC when only a truncated metric is used for solving the geodesic equation directly in RNC. 

For that purpose, we consider the previously discussed infinitesimal neighborhood of the reference point which crosses the event horizon, but now take the first curvature correction into account. Setting \pagebreak $\xi^1 = 0$ and computing the coordinate velocity of radially outward directed light rays, we find
\begin{equation}
\label{causality_geodesic_solved}
\left.v^1(\xi^0)\right|_{\xi^1=0} = \left.\frac{\mathrm d \xi^1}{\mathrm d \xi^0}\right|_{\xi^1=0} = \left(1-\frac{(\xi^0)^2}{12 M^2}\right)^{-\frac 1 2}.
\end{equation}
Making use of (\ref{2_order2}), we can show that the argument on the right-hand side of (\ref{causality_geodesic_solved}) will always remain positive. The condition on $\xi^{0\,(2)}_{\backslash \mathrm d R}$ gives, in the limit $r_0\rightarrow 2M$, $D\rightarrow 2M$, $d\rightarrow 1$, the restriction $\left|\xi^0\right|\leq 2M(\varepsilon^{1/3} + 4\varepsilon/9)$. For the square of this maximal value to be larger than $12M^2$, we would require $\varepsilon \gtrsim 1.39$ which is an invalid value for $\varepsilon$.

Relation (\ref{causality_geodesic_solved}) has the geodesics starting steeper the later they start from $\xi^1=0$ and thus already suggests that a horizon which is located at $\tau_{\text{PG}}$ will not be crossed by a geodesic starting at $\xi^0 > \tau_{\text{PG}}$. Furthermore, we find with (\ref{causality_geodesic_solved}) that $v^1(\tau_{\text{PG}})\leq \sqrt{2M/r_0}$ is only satisfied for $0<r_0\leq 2r_0$. Therefore, the lines corresponding to outward directed light rays inside the black hole are always steeper than the line corresponding to the horizon. 

We refrained from giving the full expressions for this coordinate velocity and the corresponding geodesics, since they become very lengthy. Without the restriction $\xi^1=0$ we could find the above behavior to change, however. Also, the geodesics are in fact curved lines. In order to show causality for $\xi^1\neq 0$, we therefore plot two sample geodesics in Fig.~\ref{figRNCgeod} with solid lines. We see that the steeper a line starts at $\xi^1=0$, the steeper it will continuously grow. As a consequence, lines which start later and steeper at $\xi^1=0$ compared to others will only become even steeper in comparison and outward directed light rays can indeed not leave the black hole.

If the reference point is on the event horizon $r_0=2M$ and the horizon is consequently described by the lower solid line, we see that outward directed light rays originating inside the horizon, which are given by the upper solid line and seem to be outgoing for the RNC observer, cannot escape the black hole. In fact, their distance to the horizon increases, i.e., they falls inward. Furthermore, any object outside the black hole may only interact with such light rays after it as well has crossed the horizon to the inside of the black hole.

If the expansion point is set at $r_0>2M$, however, and the horizon is therefore described by the upper solid line, light rays emitted at the reference point follow the lower solid line and diverge radially from the horizon to the outside. Both these effects can be seen in PG coordinates, too, but are not reflected in the Minkowski patch of RNC, as can be read off the dashed lines. Using solely the Minkowski metric, the distance between infalling or outward directed light rays and the horizon is constant.

The analysis in this section shows that whatever RNC patches one constructs, even those that cross an event horizon, causality is respected within the patches. This underlines the statement of normal coordinates being valid in some finite spacetime region if they are used to a finite adiabatic order.

\subsection{Outlook on other horizons}
\label{subsec: outlook}

As mentioned above, we chose the example of the eternal Schwarzschild black hole because it allowed us to analyze the behavior of normal coordinate patch sizes and of the causality within these patches in the presence of the event horizon and singularity concomitant with this geometry.

Since the normal coordinate construction works for any smooth and connected background geometry, it is reasonable to assume that our results on normal coordinate patches are generalizable to any geometry containing horizons \cite{horizons} and singularities. This includes RNC patches crossing horizons and avoiding singularities as well as the conservation of the causal structure within such patches. In detail, cosmological horizons of expanding universes surrounded by a Hubble sphere or of contracting universes (big crunch) as well as white holes would be prominent examples of such spacetimes with horizons and singularities. 

\section{Conclusion and summary}
\label{sec: conclusion}

In this article, we developed a procedure to determine the domain of validity of normal coordinate systems. Given some precision $\varepsilon$ and some set of tensors one wants to work with, we showed how to evaluate the restriction on the spacetime region. For this procedure to work and for a point to be includable in a normal coordinate neighborhood, all we required was a metric that is analytic (regular) in a region containing both the reference point (or geodesic of reference points in the case of FNC) and the point in question.

For the developed method we considered both RNC and FNCP, i.e., normal coordinates using curvature information at a point, as well as the usual FNC, i.e., normal coordinates using geometrical information along a geodesic.

The complete exponential map recapitulated in Sec.~\ref{sec: prelim} is a bijective map on the full spacetime manifold. In this article, we considered truncated versions of the exponential map, however, which cannot be bijective on the complete manifold anymore. By calculating the domain of validity for truncated normal coordinate expansions, we therefore determined the subset of the spacetime \mbox{manifold} on which the truncation of the exponential map is again sufficiently bijective (up to a precision requirement) to describe physical systems. This also manifests itself in the fact that fundamental properties of the spacetime manifold, such as causality at an event horizon, as correctly described by normal coordinates in their domain of validity.

As an example, we examined the spacetime geometry of a Schwarzschild black hole and showed in Fig.~\ref{RNCpatch} the RNC domain of validity for several different orders of the normal coordinate expansion and different precision requirements. We found that the subset of the normal coordinate reference point covered by the normal coordinates' domain of validity can in general cross the black hole's horizon, but can never include the physical singularity at its center. When translating the patch size from normal coordinates to some other, possibly global, coordinate system, the ability for this subset of the causal future to cross the horizon can only be achieved, however, in such coordinates which are regular at the horizon, such as PG coordinates. In contrast, a normal coordinate patch in Schwarzschild coordinates only reaches the horizon. Furthermore, in Fig.~\ref{RNCpatch3rd4th} we also showed in detail how an RNC patch increases in size with increasing order of the normal coordinate expansion.

The importance of normal coordinates in physical applications is due to the solutions of geodesic equations or equations of motion for arbitrary (quantum) fields possibly being unobtainable for the exact metric. Whenever this is the case, one can approximate these differential equations by using the polynomial normal coordinates and thus bring them in a systematically solvable form. This way, all possible local applications in curved spacetime can be treated perturbatively as long as one works within the domain of validity. With our findings one can now calculate this domain of validity as well as the error that stems from using a normal coordinate neighborhood compared to an exact treatment. Consequently, one can do computations which produce results one wants to compare with experimental or observational data. 

For example, the normal coordinates used in \cite{TidalBH1} to describe tidal disruption events of stars close to black holes have an external precision requirement of $10^{-4}$. Employing our procedure, we see that the calculated validity domain of the normal coordinates for this precision is large enough to accommodate the whole star. In contrast, the literature estimates the patch size as much too small. We therefore see the importance of properly calculating normal coordinate patch sizes.

\begin{acknowledgments}
It is a great pleasure to thank Marc Schneider for delightful discussions and suggestions which substantially improved the article.
\end{acknowledgments}

\appendix
\section{FNCP WITHOUT EXTERNAL INFORMATION ON TEMPORAL VALIDITY}
\label{sec: appendix a}

As mentioned in Sec.~\ref{subsec: FNCP}, if we have no external information on the temporal validity of the central geodesics Taylor expansion, the domain of validity of this coordinate system equals that of an RNC patch. The adiabatic order corresponding to this RNC patch is then given by the adiabatic order of the FNCP's orthogonal expansions.

To show this in more detail, we cannot employ the patch size computation method presented in the section on FNCP, as it allows for no restriction on $\xi^0$. Instead, we address points in the FNCP patch differently by performing two FNC expansions around the reference point as illustrated in Fig. \ref{fnc}. First, we perform the spatial RNC expansion around $p$, which is represented by the spatial geodesic $\omega_0(\kappa)$ that reaches $p'$ after some length $\kappa^0$. Second, we implement another RNC expansion, this time around $p'$ and orthogonal to $\omega_0$, i.e., using only geodesics $\Omega(\rho)$ that satisfy $\Omega(0)=p'$ and $\left.\mathrm d_\rho\Omega\right|_0\perp \left.\mathrm d_\kappa \omega_0\right|_{\kappa^0}$. We can now again determine the patch size by employing the familiar method developed for RNC.

We write the RNC of the first expansion around $p$ as usual by $\xi^{\bar\alpha}$. The second RNC of the expansion around $p'$ we denote as $\chi^{\bar\Theta}$. Notice that the vierbein ${e'}^a_\Delta(\xi)$ corresponding to the second expansion at $p'$, given in terms of the first expansion's RNC, is obtained by parallel transport along $\omega_0$. This means that, in general, we have ${e'}^a_\Delta(\xi) = {e'}^a_\Delta + {e'}^a_{\Delta,\bar{\mu}\,}\xi^{\bar\mu} + \cdots\,$, where the coefficients are evaluated at $\xi=0$, i.e., at $p$. This primed vierbein at $p$ ${e'}^a_\Delta$ is obtained from the first expansion's vierbein $e^a_\alpha$ by
\begin{equation}
\label{vierbein'_FNC_alt}
{e'}^a_1 = e^a_0, \quad {e'}^a_2, = e^a_2, \quad {e'}^a_3 = e^a_3, \quad {e'}^a_0 = \lambda^{\bar\alpha}e^a_{\bar\alpha}\;,
\end{equation}
with a residual freedom of rotating ${e'}^a_1$, ${e'}^a_2$, and ${e'}^a_3$.

For the determination of the patch size we again choose the metric and Riemann tensor as the tensors of interest and consider the metric (\ref{met_FNC_alt}) with $\xi^0=0$ up to adiabatic order 2 in the $\xi^{\bar\alpha}$. Recalling that the conditions derived from the Riemann tensor will always be more restrictive than those from the metric, we omit considering the metric in the following calculations. Similar to before, we will now employ the method for RNC patch sizes. 

To shorten the expressions, we will denote the additional $\mathcal{O}_{\text{Riem}}(\xi^2)$~terms that appear in (\ref{Riem_order2}) when computing the Riemann tensor using the truncated metric by writing $\left[R^2\right]_{\alpha\beta\gamma\delta\mu\nu}$. For the same reason, we will omit denoting terms of higher order than 2 in $\xi$ and $\chi$.

Using (\ref{Riem_order2}), the Riemann tensor at $Q$ is given by
\begin{equation}
\label{Riem_FNC_alt}
R_{\Delta\Lambda\Pi\Sigma}(\chi,\xi) = R_{\Delta\Lambda\Pi\Sigma}(\xi) + \left[R^2\right]_{\Delta\Lambda\Pi\Sigma\bar\Theta\bar\Phi}(\xi)\chi^{\bar\Theta}\chi^{\bar\Phi}\;.
\end{equation}
The $\xi$ dependence results from the $\chi$~expansion being around $p'$. We employ (\ref{Riem_order2}) once more and find for the first term on the right-hand side
\begin{eqnarray}
&& R_{\Delta\Lambda\Pi\Sigma}(\xi) = ({e'}^{a}_{\Delta}{e'}^{b}_{\Lambda}{e'}^{c}_{\Pi}{e'}^{d}_{\Sigma})(\xi) R_{abcd}(p')  \nonumber\\
&& = \left({e'}^a_\Delta  {e'}^b_\Lambda  {e'}^c_\Pi  {e'}^d_\Sigma \right)\!(\xi)\, e^\alpha_a e^\beta_b e^\gamma_c e^\delta_d \nonumber \\ 
&& \phantom{=}\, \times \left( R_{\alpha\beta\gamma\delta} + \left[R^2\right]_{\alpha\beta\gamma\delta\bar\mu\bar\nu}\xi^{\bar\mu}\xi^{\bar\nu}\right)  \nonumber\\
&& = \left({e'}^{a}_{\Delta}{e'}^{b}_{\Lambda}{e'}^{c}_{\Pi}{e'}^{d}_{\Sigma}\right)\!(\xi) \nonumber\\
\label{Riem_FNC_alt1}
&&\phantom{=}\, \times \left( R_{abcd} + \left[R^2\right]_{abcdmn}e^m_{\bar\mu}e^n_{\bar{\nu}\,}\xi^{\bar\mu}\xi^{\bar\nu} \right).
\end{eqnarray}
We obtain the second term on the right-hand side by simple tensor transformation
\begin{eqnarray}
&& \left[R^2\right]_{\Delta\Lambda\Pi\Sigma\bar\Theta\bar\Phi}(\xi)\chi^{\bar\Theta}\chi^{\bar\Phi}  \nonumber \\
\label{Riem_FNC_alt2}
&& = \left({e'}^{a}_{\Delta}{e'}^{b}_{\Lambda}{e'}^{c}_{\Pi}{e'}^{d}_{\Sigma}\right)\!(\xi)\left[R^2\right]_{abcdmn}{e'}^m_{\bar\Theta} {e'}^n_{\bar\Phi}\chi^{\bar\Theta}\chi^{\bar\Phi}\,,
\end{eqnarray}
where we dropped the $\xi$ dependence of ${e'}^m_{\bar\Theta} {e'}^n_{\bar\Phi}$ as we wish to only consider terms up to order 2 in $\xi$ and $\chi$. In contrast, we kept the $\xi$ dependence of the vierbein factor in front because we want to be able to compare to (\ref{Riem_FNC_alt1}). 

Plugging (\ref{Riem_FNC_alt1}) and (\ref{Riem_FNC_alt2}) back into (\ref{Riem_FNC_alt}), we find
\begin{eqnarray}
&& R_{\Delta\Lambda\Pi\Sigma}(\chi,\xi) = \left({e'}^{a}_{\Delta}{e'}^{b}_{\Lambda}{e'}^{c}_{\Pi}{e'}^{d}_{\Sigma}\right)\!(\xi) \left( R_{abcd} \phantom{\chi^{\bar\Theta}}\right.  \nonumber \\
\label{Riem_FNC_alt_order2}
&& + \left.\left[R^2\right]_{abcd\bar\mu\bar\nu}\xi^{\bar\mu}\xi^{\bar\nu} + \left[R^2\right]_{abcd\bar\Theta\bar\Phi}\chi^{\bar\Theta}\chi^{\bar\Phi} \right).
\end{eqnarray}
Here we must notice that $({e'}^{a}_{\Delta}{e'}^{b}_{\Lambda}{e'}^{c}_{\Pi}{e'}^{d}_{\Sigma})(\xi)R_{abcd}$ is the correct term in adiabatic order $0$ since it contains no geometrical information at $p'$, but only at $p$, and is tensor transformed using the correct vierbein ${e'}^a_\Delta(\xi)$ at $p'$. The second and third terms are mismatch terms of very much the same structure as the one in (\ref{Riem_order2}). We denoted the combination of vierbein and Riemann tensors as $\left[R^2\right]_{abcdmn}{e'}^m_{\bar\mu} {e'}^n_{\bar\nu} = \left[R^2\right]_{abcd\bar\mu\bar\nu}$. For the same reason of neglecting terms of higher order in any of the two RNC as below (\ref{Riem_FNC_alt2}), any tensor transformation to or from the $\chi$ applied here can only use the primed vierbein ${e'}^a_\Delta$ at~$p$.

In order for the mismatch term in (\ref{Riem_FNC_alt_order2}) to be negligibly small compared to the correct term, we require that
\begin{equation}
\label{FNC_alt_cond1}
\left|\left[R^2\right]_{abcd\bar\mu\bar\nu}\xi^{\bar\mu}\xi^{\bar\nu} + \left[R^2\right]_{abcd\bar\Theta\bar\Phi}\chi^{\bar\Theta}\chi^{\bar\Phi}\right| = \varepsilon\left|R_{abcd}\right|\,.
\end{equation}
This can then be tensor transformed to the $\xi$ or $\chi$ using either $e^a_\alpha$ or ${e'}^a_\Delta$ (the latter only at $p$) to give a condition that is related to (\ref{4)_general_Riem_1}). This relation is the same as that between the condition for points $(x,y)$ within a circle of radius $L$ around the origin ($x^2 + y^2 \leq L^2$) and the condition for points $x$ on a line segment of length $2L$ symmetric around the origin ($|x|\leq L$).

The usual Taylor expansion of the Riemann tensor in this setup of RNC can be obtained as follows (we will again not denote higher order terms). In the $\chi$, the Taylor series of the Riemann tensor around $p'$ reads $R_{\Delta\Lambda\Pi\Sigma}(\chi,\xi) = R_{\Delta\Lambda\Pi\Sigma}(\xi) + R_{\Delta\Lambda\Pi\Sigma,\bar\Theta}(\xi)\chi^{\bar\Theta}$. Plugging into this expression the Taylor expansion of the Riemann tensor around $p$ in $\xi$ $R_{\alpha\beta\gamma\delta}(\xi) = R_{\alpha\beta\gamma\delta} + R_{\alpha\beta\gamma\delta,\vec\mu}\xi^{\vec\mu}$ and ignoring all terms of higher adiabatic order than 1, we find the series expansion
\begin{eqnarray}
&& R_{\Delta\Lambda\Pi\Sigma}(\chi,\xi) = \left({e'}^{a}_{\Delta}{e'}^{b}_{\Lambda}{e'}^{c}_{\Pi}{e'}^{d}_{\Sigma}\right)\!(\xi) \left( R_{abcd} \phantom{\chi^{\bar\Theta}}\right.  \nonumber \\
\label{Riem_FNC_alt_Tay}
&& + \left. R_{abcd,\bar\Theta}(\xi)\chi^{\bar\Theta} + R_{abcd,\bar{\mu}}\xi^{\bar\mu}\right).
\end{eqnarray}

We demand agreement between (\ref{Riem_FNC_alt_order2}) and (\ref{Riem_FNC_alt_Tay}), thus requiring the first order term in the latter to be negligible, and obtain the condition
\begin{equation}
\label{FNC_alt_cond2}
\left|R_{abcd,\bar\Theta}(\xi)\chi^{\bar\Theta} + R_{abcd,\bar\mu}\xi^{\bar\mu}\right| = \varepsilon \left|R_{abcd}\right|\,.
\end{equation}
We can again tensor transform this condition to the $\xi$ or $\chi$, using the respective vierbein only at $p$, and obtain conditions which are analogous to (\ref{4)_general_Riem_2}) in the same way (\ref{FNC_alt_cond1}) was to (\ref{4)_general_Riem_1}).

Alternatively, when calculating the patch size determined by (\ref{4)_general_Riem_1}) and (\ref{4)_general_Riem_2}) in any other $x$ coordinate system according to Step~5 of the general method, we will have to transform the Riemann components to those coordinates and find patch size restrictions on the RNC in terms of these components. These conditions will then have the same structure as the two terms in (\ref{FNC_alt_cond1}) and (\ref{FNC_alt_cond2}), respectively.

The close relation between conditions (\ref{4)_general_Riem_1}) and (\ref{4)_general_Riem_2}) with (\ref{FNC_alt_cond1}) and (\ref{FNC_alt_cond2}) shows that the patch sizes of RNC and FNCP determined using the Riemann tensor are in fact the same. In particular, in the limit of $\xi^{\bar\alpha}\rightarrow 0$ the geodesic $\Omega(\rho)$ follows $\gamma(\xi^0)$ arbitrarily closely, such that the domain of validity of the $\chi$~RNC patch describes the $\xi^0$~restrictions arbitrarily well. If we did not set $\xi^0 =0$ above, we would find the additional restriction ${\dot R}_{\alpha\bar\mu\beta\bar\nu}\xi^0 \ll R_{\alpha\bar\mu\beta\bar\nu}$, which is just the analogue to (\ref{4)_general_Riem_2}).

This procedure is readily generalized to higher orders and in these cases also yields the expected $\xi^0$~restrictions $\ddot{R}_{\alpha\bar\mu\beta\bar\nu}(\xi^0)^2 \ll R_{\alpha\bar\mu\beta\bar\nu}$ and so on.

As we discussed above, this result exactly coincides with our geometrical intuition, and we therefore expect the patch size equality to be generalizable for all tensors and to not be a unique feature of the conditions derived from the Riemann tensor.

\section{FIRST DERIVATIVES OF THE RIEMANN TENSOR IN RNC}
\label{sec: appendix b}
The explicit expressions for the derivatives of the Riemann tensor in RNC are obtained from (\ref{Riem_RNC_deriv}) and read as follows:
\begin{eqnarray}
\label{dRiem_RNC}
&& R_{0110,0} = R_{2323,0} = \frac{6M}{{r_0}^4} \sqrt{\frac{2M}{r_0}}\;, \nonumber \\
&&    R_{2020,0} = R_{1221,0} = R_{1021,2} = R_{3032,2} = \frac{3M}{{r_0}^4} \sqrt{\frac{2M}{r_0}}\;, \nonumber \\
&& R_{1010,1} = R_{2332,1} = \frac{6M}{{r_0}^4} \;,\nonumber \\
&& R_{0220,1} = R_{1212,1} =   R_{0120,2} = R_{2331,2}  = \frac{3M}{{r_0}^4}\;,
\end{eqnarray}
where $r_0$ is the radial value of the RNC reference point. Components with an index interchange $2 \leftrightarrow 3$ remain unchanged due to the symmetry between the $\xi^2$ and $\xi^3$~direction.

	
	%
	
	
	
	
	\bibliography{library.bib}
	
\end{document}